\documentclass[twocolumn,journal]{IEEEtran}
\usepackage{framed}
\usepackage{color}
\usepackage[makeroom]{cancel}
\usepackage{rotating}
\usepackage{makeidx}

\usepackage{graphicx,subfigure}
\usepackage{caption}
\usepackage[cmex10]{amsmath}
\newcounter{subeqn} \renewcommand{\thesubeqn}{\theequation\alph{subeqn}}%
\makeatletter
\@addtoreset{subeqn}{equation}
\makeatother
\newcommand{\subeqn}{%
	\refstepcounter{subeqn}
	\tag{\thesubeqn}
}

\usepackage{algorithm}
\usepackage{amssymb}
\usepackage{mathrsfs}
\usepackage{balance}
\usepackage{enumitem}
\usepackage{hyperref}
\usepackage[noadjust]{cite}

\newtheorem{proposition}{Proposition}

\newtheorem{remark}{Remark}

\newcommand{\bs}{\boldsymbol}
\newcommand{\Tr}{\mathrm{Tr}}

\newcommand{\bts}{\begingroup\textstyle}
\newcommand{\ets}{\endgroup}

\begin{document}
	\title{Full-Duplex Enabled Mobile Edge Caching: From Distributed to Cooperative Caching}

	
	\author{\IEEEauthorblockN{Thang~X.~Vu, \IEEEmembership{Member, IEEE}, Symeon~Chatzinotas, \IEEEmembership{Senior Member, IEEE}, Bj\"orn~Ottersten, \IEEEmembership{Fellow, IEEE}, and Anh~Vu~Trinh
		}\\
	\thanks{This research is supported, in part, by the ERC AGNOSTIC project under code R-AGR-3283, the FNR CORE ProCAST project under code R-AGR-3415-10, and Vietnam National University, Hanoi (VNU), under Project No. QG.18.39. Parts of this work was presented to the conference IEEE WCNC 2019 \cite{Vuwcnc19}.}
	\thanks{T.~X.~Vu, S. Chatzinotas and B. Ottersten are with the Interdisciplinary Centre for Security, Reliability and Trust (SnT) -- University of Luxembourg, L-1855 Luxembourg. Email: \{thang.vu, symeon.chatzinotas, bjorn.ottersten\}@uni.lu.}
	\thanks{A.~V.~Trinh is with the Department of Electronics and Telecommunications -- VNU University of Engineering and Technology, Hanoi, Vietnam. Email: vuta@vnu.edu.vn}
	\thanks{This work is accepted to the \emph{IEEE Transactions on Wireless Communications}. Personal use is permitted, but republication/redistribution requires IEEE permission.}
	}
	
	\providecommand{\keywords}[1]{\textbf{\textit{Index terms---}} #1}
	
	\date{}
	
	\maketitle
	\thispagestyle{plain}
	\begin{abstract}
Mobile edge caching (MEC) has received much attention as a promising technique to overcome the stringent latency and data hungry requirements in future generation wireless networks. Meanwhile, full-duplex (FD) transmission can potentially double the spectral efficiency by allowing a node to receive and transmit in the same time/frequency block simultaneously. In this paper, we investigate the delivery time performance of full-duplex enabled MEC (FD-MEC) systems, in which the users are served by distributed edge nodes (ENs), which operate in FD mode and are equipped with a limited storage memory. Firstly, we analyse the FD-MEC with different levels of cooperation among the ENs and take into account a realistic model of self-interference cancellation. {Secondly, we propose a framework to minimize the system delivery time of FD-MEC under both linear and optimal precoding designs}. Thirdly, to deal with the non-convexity of the formulated problems, two iterative optimization algorithms are proposed based on the inner approximation method, whose convergence is analytically guaranteed. Finally, the effectiveness of the proposed designs are demonstrated via extensive numerical results. {It is shown that the cooperative scheme mitigates inter-user and self interference significantly better than the distributed scheme at an expense of inter-EN cooperation. In addition, we show that minimum mean square error (MMSE)-based precoding design achieves the best performance-complexity trade-off, compared with the zero-forcing and optimal designs.}  
\end{abstract}

\section{Introduction}
Among potential enabling technologies to tackle with stringent latency and data hungry requirements in future wireless networks, mobile edge caching (MEC) has received much attention. The basic premise of MEC is to bring the content close to end users via distributed storages through out the network. Caching usually comprises a placement phase and a delivery phase. In the former, which is implemented during off-peak periods when the network resources are abundant, popular content is prefetched in the distributed caches. The latter usually occurs during peak-hours when the content requests are revealed. If the requested content is available in the edge node's local storage, it can be served directly without being sent from the core network. In this manner, MEC enables  significant reduction in transmission latency and backhaul traffic thus mitigating network congestion \cite{Borst2010}. 
Joint design for content caching and physical layer transmission has attracted much attention recently. The main idea is to take into account the cached content at the edge nodes when designing the signal transmission to reduce costs on both access and backhaul links. Since some requested files are available in the edge node's cache, proper design is required for content selection combined with broad/multi-cast transmission design to improve the system performance, including energy efficiency (EE) \cite{Gabry2016, Vu18TWC,Vu18WCL,Liu2016} and content delivery time  \cite{Tran2016,Tao17:NDT,Sengupta2016b}. The role of caching in wireless device-to-device (D2D) networks is analysed in \cite{ZhaXiaWuLi2016,Gregori2016,Ji2016}.  The performance of cache-aided wireless networks can be further improved by joint optimization of caching along with routing and resource allocation \cite{Khreishah2016}. It is worth noting that these works study the caching in the half-duplex (HD) systems.

Meanwhile, full-duplex (FD) has shown great potential as the transmission technique to overcome the spectral scarceness in next generation wireless networks by allowing a node to transmit and receive in the same time/frequency resource \cite{Sabha14:FullDuplex,Lei18_FDNOMA,Tan2,S1,S2}. The foreseen benefit of FD is, however, not without limitation. The major issue lies in the interference caused by the FD transmissions. In fact, a few FD links might result in continuous interference towards neighbouring nodes, in addition to the self interference. Fortunately, thanks to recent developments in the self-interference cancellation, FD can potentially double the spectral efficiency compared with the conventional HD counterpart \cite{Sabha14:FullDuplex}. The employment of FD systems with caching capability has the potential to further improve the system performance. 


\subsection{Related works}
Despite that cache-aided HD has been well studied in the literature, the investigation on cache-aided FD systems is limited. 
The authors in \cite{Marso17:CacheFD} show that cache-aided FD small cell networks can provide cache hit enhancements compared with the HD system. In that work, by modelling the base stations and users as a coupled Poison Point Process (PPP) with the edge nodes, coverage probability and successful delivery rates are analysed. The role of caching in FD D2D networks is investigated in \cite{Nasl:FD_D2D_2018,Hema:FD_D2D:18} via stochastic geometry analysis. By considering all possible operating modes of an arbitrary device, the success probability is derived in \cite{Nasl:FD_D2D_2018} as a function of the caching capacity and interference distribution.  It is shown in \cite{Hema:FD_D2D:18} that allowing a hybrid deployment of FD and HD modes can potentially further improve the coverage probability in cluster-based FD networks. The authors of \cite{Vu19:CL} derive closed-form expression for the successful delivery probability of a cached-aided FD system by taking into account the distribution of all wireless links. The worst case normalized delivery time (NDT) in heterogeneous networks  is studied in \cite{Kakar18:NDT_FullDuplex} with FD relaying nodes. However, the results in \cite{Kakar18:NDT_FullDuplex} are based on an optimistic assumption of perfect self-interference cancellation. In practice, there always remains residual interference after the self-interference cancellation \cite{Knox12:SAFD,Bharadia14:FDM}.

\subsection{Our contributions}
In this paper, we study the performance of FD-enabled MEC (FD-MEC) systems, in which the users demand content via distributed cache-assisted edge nodes (ENs). The ENs operate in FD mode and connect to the core network via wireless backhauls. Unlike previous works on cache-assisted HD systems \cite{Vu18TWC}, the FD transmissions can cause significant self-interference, in addition to inter-user interference. Our goal is to minimize the system delivery time via a joint design of precoding vectors on both backhaul and access links, by taking into consideration the cached content and interference patterns. The contributions of this paper are as follows:
\begin{itemize}
	\item Firstly, we investigate the delivery time performance of FD-MEC systems by considering a realistic model of the self-interference cancellation. Our work is fundamentally different from \cite{Marso17:CacheFD,Nasl:FD_D2D_2018,Hema:FD_D2D:18,Vu19:CL} in terms of performance metric and analysis method. Compared with \cite{Kakar18:NDT_FullDuplex}, which understands the cache-aided FD system from the information-theoretic asymptotic aspect and relies on the perfect assumption of self-interference cancellation, we consider the practical interference cancellation model and {focus on precoding vectors design}.
	\item Secondly, we analyse the system via two network architectures for different levels of cooperation among the ENs, namely distributed and cooperative caching. {For each architecture, an optimization problem is formulated that minimizes the system delivery time based on both linear and optimal (non-linear) precoding designs. The formulated problems optimize the precoding vectors while minimizing both inter-user and self interference.}
	\item Thirdly, to cope with the non-convexity of the formulated problems caused by the self-interference, we propose two iterative optimization algorithms based on the inner approximation method. The convergence of the proposed iterative algorithms are analytically guaranteed. 
	\item Finally, extensive numerical results are presented to demonstrate the effectiveness of the proposed algorithms and the benefit of the FD-MEC over the half-duplex system in certain scenarios. 
\end{itemize}

The rest of this paper is organised as follows. Section~\ref{sec:SysModel} presents the system model and the caching strategies. Section~\ref{sec:SigModel} gives the signal transmission details of the two caching modes. Section~\ref{sec:Dist} proposes the precoding vectors design for the distributed caching scheme. Section~\ref{sec:coop} optimizes the precoding vectors for the cooperative caching scheme. Numerical results are shown in Section~\ref{sec:results}. Finally, Section~\ref{sec:conclusion} provides conclusions and discussions. 

\emph{Notation:} $(.)^H, (.)^T$ and $(.)^{-1}$ denote the conjugate operator, transpose operator, and the inverse matrix, respectively. $\mathrm{Tr}(\bs{X})$ denotes the trace of matrix $\bs{X}$.

\section{System and caching model}\label{sec:SysModel}
\begin{figure*}
	\normalsize
	\centering
	\subfigure[Distributed architecture]{\includegraphics[width=\columnwidth]{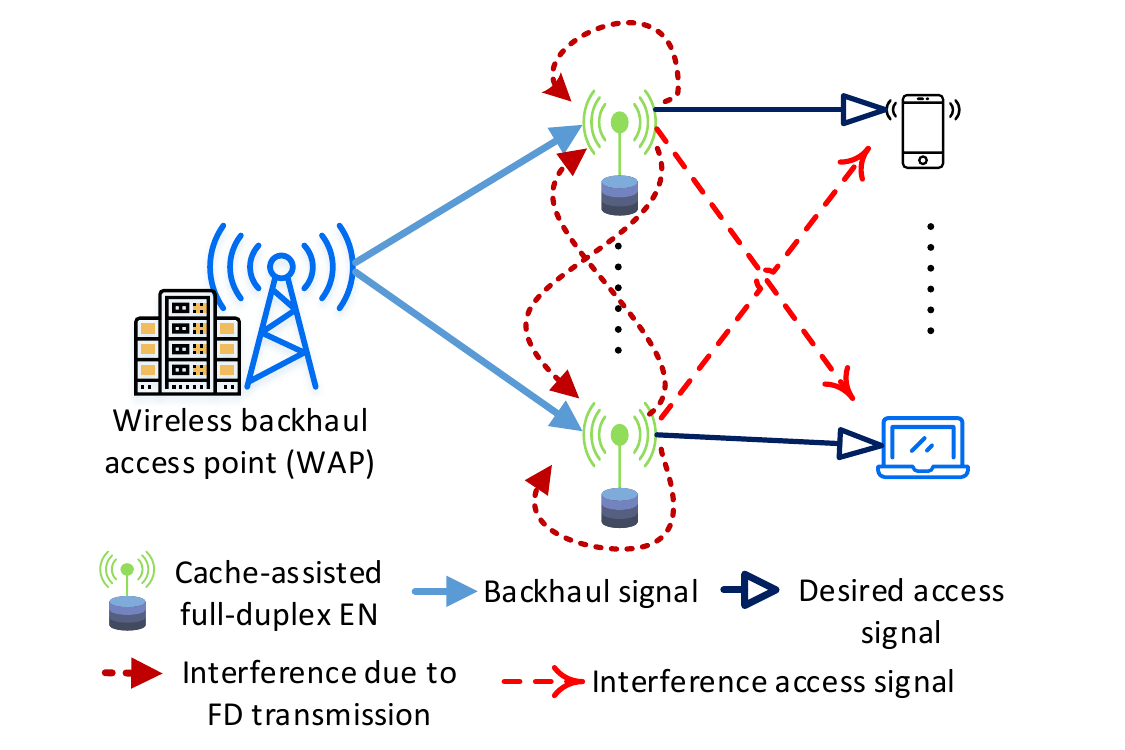}}	
	\subfigure[Cooperative architecture]{\includegraphics[width=\columnwidth]{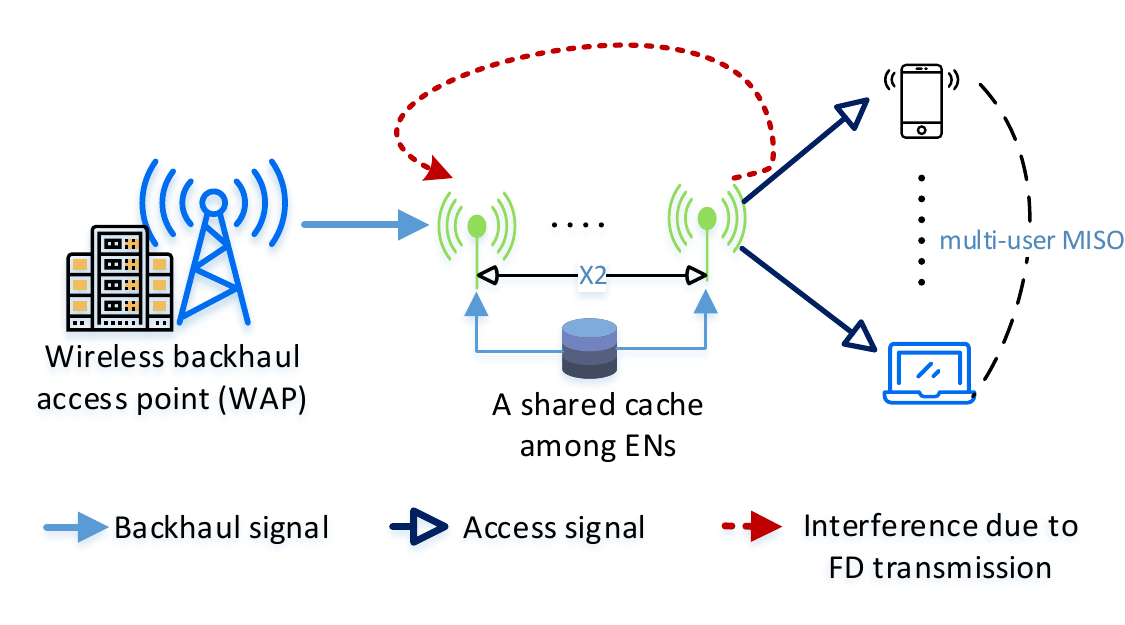}}	
	\caption{Block diagram of FD-MEC. In the distributed caching architecture (a), each EN has its own cache and serves its intended user separately, which imposes three types of interferences at the users and ENs's (backhaul) receiver. In the cooperative caching architecture (b), the ENs share a common cache and cooperatively serve the users via joint access transmission. In both architectures, the ENs decode the backhaul signal separately.} \label{fig:SysModel}
\end{figure*}
We consider a cache-aided FD system, in which the users are served via a number of distributed ENs, e.g., pico- and small cell base stations, as depicted in Fig.~\ref{fig:SysModel}. The ENs operate in FD mode and connect to the core network via a wireless backhaul access point (WAP), e.g., macro base station. The users can only access data from the ENs via wireless access channels, i.e., there is no direct link between the users and the WAP. The WAP is equipped with $N$ antennas, while the ENs and users are equipped with a single-antenna. {Let $K$ denote the number of ENs with $K \leq N$. It is assumed that each EN serves only one user at a time \cite{Marso17:CacheFD}, thus the number of active users is also $K$}\footnote{When the number of users is greater than $K$, the EN can serves its active users via, e.g., time division multiplexing. Studying such scenario is beyond the scope of this paper.}. The WAP is assumed to have access to a library of $F$ contents, denoted by $\mathcal{F} = \{f_1, \dots, f_F\}$. Without loss of generality, all content is assumed to have equal size of $Q$ bits. To leverage the backhaul during peak-hours, the EN is equipped with a storage memory of $MQ$ bits, where $M < F$.

We consider two network architectures depending on the level of collaboration among the ENs: i)  \emph{Distributed caching - separate access transmission} (DCST) and ii) \emph{Cooperative caching with joint access transmission} (CCJT). 
\subsection{Distributed caching - separate transmission (DCST) mode}
The DCST mode does not require any collaboration among the ENs, hence minimizing the system's signalling overhead (Fig.~\ref{fig:SysModel}a). In this mode, each EN stores the content in its local cache, and operates independently from other ENs. {The benefit of this mode is the scalability and flexibility. However, since the transmissions of the ENs are independent, severe inter-user interference can occur, resulting in a significant performance degradation}. Details on this mode will be presented in the next section.
\subsection{Cooperative caching - joint transmission (CCJT) mode}
In this mode, all the ENs share a common cache and cooperatively serve the user requests (Fig.~\ref{fig:SysModel}b). Such cooperation can be enabled via the dedicated X2 link \cite{X2}. {Since the ENs jointly transmit the requested contents to the users, inter-user interference can be efficiently mitigated, thus the system performance can be largely improved compared to the DCST mode. These improvements, however, require inter-EN collaboration and additional signalling overhead}. It is worth noting that although the inter-user interference can be avoided, self-interference still exits at the ENs' receivers due to the ENs' FD transmission. 
\begin{remark}
	Intuitively, these two modes serve as the two extremes of the network architecture when inter-EN cooperation is allowed. Analysing these two modes provide the lower- and upper-bound for the performance of FD-MEC systems. 	
\end{remark}
\subsection{Content popularity and caching model}
We consider the most popular content popularity model, i.e., the Zipf distribution \cite{Breslau:Zipf:1999}. The probability for file $f_n$ being requested is equal to 
\begin{align}
\nu_n = \frac{n^{-\xi}}{\sum_{m=1}^F m^{-\xi}} ,\label{eq:qn}
\end{align}
where $\xi$ is the Zipf skewness factor. 

{This paper focuses on off-line caching delivery phase, in which the \emph{content placement phase} is predetermined and executed during off-peak times \cite{TaoCheZhoYu2016,Vu18TWC,Vu18WCL,Liu2016,ZhaXiaWuLi2016,Ji2016}}. We consider a generic caching policy (cache placement) $\boldsymbol{\mu} = [\mu_1, \dots, \mu_F]$, where $0 \le \mu_n \le 1$ denotes portions of file $f_n$ cached at the ENs. In order to meet the memory constraint, it must hold that $\sum_{n=1}^F \mu_n\leq M$. The motivation behind the generic caching policy is that it allows studying different caching strategies. In the  most popular caching, we have
$\bs{\mu}_{\rm Pop} = [\underbrace{1, \dots, 1}_{\times M}, 0, \dots, 0]$.
\section{Signal transmission model} \label{sec:SigModel}
In this section, we provide details on the signal transmission in two DCST and CCJT modes. Since each EN serves only one user, we use the user index and EN index interchangeably, e.g., EN $k$ means the EN serving user $k$. When user $k$ demands a file, it sends the requested file index $d_k$ to its serving EN $k$. The EN $k$ first checks its local cache. If (portions of) the requested content is available in the cache, it serves the user directly. Otherwise, the EN $k$ will ask for the non-cached parts from the WAP via the wireless backhaul before serving the user. 

Denote $\mathcal{U}_{C}$ as the set of ENs which have only some portions of the requested files in their caches, i.e., $\mathcal{U}_C = \{k~ |~ \mu_{d_k} < 1\}$. Let $K_C = |\mathcal{U}_C|$. Without loss of generality, \emph{we assume the first $K_C$ ENs cache only parts of the requested files} for ease of presentation, i.e., $\mathcal{U}_C = \{1, 2, \dots, K_C\}$. Because each EN serves only one user, we also refer $\mathcal{U}_C$ as the set of the users served by the ENs in $\mathcal{U}_C$.  This way, any EN $l \notin \mathcal{U}_C$, i.e., $K_C < l \le K$, has the whole  requested file in its cache. 

The channel fading coefficients, including the path loss, of all the links are defined in Table~\ref{tab:chan}. Full channel state information is assumed to be known at the transmitter sides. 

\subsection{Signal transmission in DCST mode}
\begin{table}[b]
	\centering
	\caption{\textsc{Channel fading coefficients, including the pathloss}}\label{tab:chan}
	\begin{tabular}{|l || l |}
		\hline 
		\bfseries{Notation} &\textbf{Explanation}. \\
		\hline 
		$\bs{g}_k \in \mathbb{C}^{1\times N}$ & WAP $\to$ EN $k$ backhaul channel coefficients \\
		$f_{kl} \in \mathbb{C}$ & EN $l \to $ EN $k$ inter-EN channel coefficients due to\\
		~& the FD transmission\\
		$f_{kk} \in \mathbb{C}$ & Self-interference at EN $k$ due to the FD transmission\\
		$h_{kl} \in \mathbb{C}$ & EN $l\to$ user $k$ access channel coefficient \\
		$\bs{h}_k \in \mathbb{C}^{1\times K}$ & $[h_{k1}, h_{k2}, \dots, h_{kK}]$ - channel coefficients from \\
		~& all ENs to user $k$  \\
		\hline 
	\end{tabular}
\end{table}

\subsubsection{Signal transmission on backhaul links}

Since the ENs not in $\mathcal{U}_C$ have the whole requested files in their cache, the WAP only sends the non-cached parts of the requested files to the ENs in $\mathcal{U}_C$ via the backhaul. 
Let $s_{E,k}, \forall k\in\mathcal{U}_C,$ denote the data symbol target to EN $k$ from the WAP. The WAP first precodes the data before sending on the backhaul. In this paper, we consider a linear minimum mean square error (MMSE) precoding for the backhaul transmission. In this design, the beamforming vector for EN $k \in\mathcal{U}_C$ is given as $\bs{w}_k = \sqrt{q_k}\tilde{\bs{w}}_k$, where $q_k$ is the power factor allocated to EN $k$ and $\tilde{\bs{w}}_k$ is the MMSE beamforming vector that is the $k$-th column of the MMSE beamforming matrix $\bs{G}^H_C (\bs{G}_C \bs{G}^H_C + \sigma^2I)^{-1}$, where $\bs{G}_C$ is the channel matrix from the WAP's antennas to the ENs in $\mathcal{U}_C$, i.e., $\bs{G}_C = [\bs{g}^T_1, \dots, \bs{g}^T_{K_C}]^T$. The size of $\bs{G}_C$ is $K_C \times L$.

The received signal at EN $k \in \mathcal{U}_C$, $y_{E,k}$, is given as
\begin{align}
		y_{\mathrm{E},k} =&~ \bs{g}_k^H \bs{w}_k s_{\mathrm{E},k}  +
{\sum}_{k\neq l\in \mathcal{U}_C}\bs{g}^H_k \bs{w}_l s_{\mathrm{E},l} \label{eq:yE}\\ 
& +\sqrt{p_k}f_{kk}s_{\mathrm{U},k}
+ {\sum}_{k\neq l=1}^K\sqrt{p_l}f_{kl}s_{\mathrm{U},l} +  n_{\mathrm{E},k},  \notag
\end{align}
where $s_{U,k}$ is the transmit symbol from EN $k$ to user $k$. In \eqref{eq:yE}, the first term is the intended signal for EN $k$; the second term  represents the inter-EN interference on the backhaul channels; the third term represents the self-interference at EN $k$ due to the FD transmission; the fourth term is interference at EN $k$ due to the FD transmission of other ENs; and $n_{E,k}$ is the Gaussian noise with zero mean and variance $\sigma^2$. 

In order to decode $y_{E,k}$, EN $k$ performs interference cancellation to mitigate the self interference, since $s_{U,k}$ is known. After the interference cancellation, there remains a residual interference with power $\eta p_k$, where $\eta$ is a Gamma distributed random variable \cite{FD_Gamma2016} representing the self-interference cancellation efficiency with a mean $\bar{\eta}$. The common value of $\bar{\eta}$ is less than $-40$\textit{dB} depending on the hardware and interference cancellation techniques \cite{Knox12:SAFD,Bharadia14:FDM}. We note that although the self interference can be effectively eliminated, there remain two interfering signals (the second and fourth terms in \eqref{eq:yE}). By treating interference as noise, the backhaul achievable information rate for EN $k \in \mathcal{U}_C$ is given as
\begin{align}
	C_{\mathtt{dist},k} =& W\log\Big(1 + \frac{|\bs{g}_k^H\tilde{\bs{w}}_k|^2 q_k}{I_{\mathtt{dist},k} + \sigma^2}\Big),  \label{eq:Ck dis}
\end{align}
where $W$ is the channel bandwidth and $I_{\mathtt{dist},k} = \sum_{k\neq l \in \mathcal{U}_C} |\bs{g}_k^H \tilde{\bs{w}}_l|^2q_l + \eta p_k + \sum_{k\neq i=1}^Kp_i|f_{ki}|^2$ is the total interference at EN $k$.

The total transmit power at the WAP is $P_{BS} = \sum_{k\in \mathcal{U}_C} {\parallel\tilde{\bs{w}}_k\parallel}^2 q_k = ~\sum_{k=1}^{K_C} {\parallel\tilde{\bs{w}}_k\parallel}^2 q_k$.
\subsubsection{Signal transmission on access links}
In the distributed caching architecture, each EN serves its user independently. Let $s_{U,k}$ denote the signal sent from EN $k$ to user $k$. The received signal at user $k$, $\forall k$, is given as
\begin{align}
y_{U,k} = \sqrt{p_k}h_{kk}s_{U,k} + {\sum}_{k\neq i = 1}^K \sqrt{p_i}h_{ki}s_{U,i} + n_{U,k}, \label{eq:yU}
\end{align}
where $h_{ki}$ is defined in Table~\ref{tab:chan} and $n_{U,k}$ is the Gaussian noise with zero mean and variance $\sigma^2$. The second term in \eqref{eq:yU} represents the inter-user interference on the access links caused by the transmission of other ENs. 

By treating interference as noise, the achievable information rate for user $k$ (on the access links) is 
\begin{align}\label{eq:Rk dis}
R_{\mathtt{dist},k} =& W\log\Big(1 + \frac{p_k|h_{kk}|^2}{\sum^K_{k\neq i = 1}p_i|h_{ki}|^2 +  \sigma^2}\Big).
\end{align}

\subsection{Signal transmission in CCJT mode}

\subsubsection{Signal transmission on backhaul links}
The backhaul transmission in CCJT mode is similar to the one in DCST mode. However, the access transmission in CCJT mode is different from DCST mode, since the ENs jointly serve the users. Let $x_k$ denote the transmit signal from EN $k$ to user $k$ (on the access links), whose details will be presented in Sec. \ref{sec:sig cop}. 
The received backhaul signal at EN $k \in \mathcal{U}_C$ is given as follows:
\begin{align}
y_{\mathrm{E},k} =&~ \begingroup\textstyle\bs{g}_k^H \bs{w}_k s_{\mathrm{E},k}  + n_{E,k} \endgroup\notag
\\
&+ \begingroup\textstyle \underbrace{{\sum}_{k\neq l \in \mathcal{U}_C}   \bs{g}^H_k \bs{w}_l s_{\mathrm{E},l}}_{(a)} \endgroup
+ \underbrace{{\sum}_{i=1}^Kf_{ki}x_i}_{(b)}, \label{eq:yE1} 
\end{align}
where $(a)$ is the interference on the backhaul links, $(b)$ is the interference due to the FD transmission of all the ENs, and $n_{E,k}$ is the Gaussian noise with zero mean and variance $\sigma^2$. 

In the cooperative mode, the precoding vectors and transmitted data are shared among all the ENs. Therefore, $x_k, \forall k$ is known at every EN. In order to decode $y_{E,k}$, the EN $k$ first performs interference cancellation on the aggregated interference $(b)$. After self-interference cancellation, there remains a residual interference with power $\eta \sum_{l=1}^K |x_l|^2$, where $\eta$ is the self-interference cancellation efficiency \cite{Bharadia14:FDM, FD_Gamma2016} which is modelled as a Gamma distributed random variable with a mean $\bar{\eta}$.
By treating interference as noise, the backhaul achievable information rate for EN $k$, $\forall k \in \mathcal{U}_C$, is given as
\begin{align}
C_{\mathtt{coop},k} =& W\log\Big(1 + \frac{|\bs{g}^H_k\tilde{\bs{w}}_k|^2 q_k}{I_{\mathtt{coop},k} + \sigma^2}\Big), \label{eq:Ck cop}
\end{align}
where $I_{\mathtt{coop},k} = \sum_{k\neq l\in \mathcal{U}_C} |\bs{g}_k^H \tilde{\bs{w}}_l|^2q_l  + \eta \sum_{l=1}^K |x_l|^2$. 

\subsubsection{Signal transmission on access links}\label{sec:sig cop}
In CCJT mode, the ENs serve all users in a cooperative manner. Therefore, the access links can be seen as a multi-user MISO channel $\bs{H} = [\bs{h}^T_1, \dots, \bs{h}^T_K]^T$, where $\bs{h}_k = [h_{k1}, \dots, h_{kK}]$ is the channel fading vector from all the ENs to user $k$. The ENs first jointly precode the data before transmitting to the users. 

Let $\bs{v}_k \in \mathbb{C}^{K\times 1}$ denote a generic precoding vector for user $k$, and $v_k[l]$ denote the $l$-th element of $\bs{v}_k$. The transmit signal at EN $k$ is $x_k = \sum_{l=1}^K v_l[k] s_{U,l}$, where $s_{U,l}$ is the data symbol dedicated to user $l$. 

The received signal at user $k$ in the CCJT mode is 
\begin{align}
y_{U,k} =& \sum_{l=1}^K h_{kl} x_l + n_{U,k} = \sum_{l=1}^K h_{kl} \sum_{i=1}^K v_i[l] s_{U,i} + n_{U,k} \notag \\
=&\bs{h}_k \bs{v}_k s_{U,k} + {\sum}_{i\neq k}\bs{h}_k \bs{v}_i s_{U,i} + n_{U,k}, \label{eq:yU2}
\end{align}
where the first term is the desired signal, the second term is the aggregated inter-user interference, and $n_{U,k}$ is the Gaussian noise with zero mean and variance $\sigma^2$.

By treating interference as noise, the achievable information rate (on the access link) for user $k$ is 
\begin{align}\label{eq:Rk cop}
R_{\mathtt{coop},k} =& W\log\Big(1 + \frac{|\bs{h}_k\bs{v}_k|^2}{\sum_{l\neq k}|\bs{h}_k\bs{v}_l|^2 + \sigma^2}\Big).
\end{align}
%

\begin{remark}
Although the DCST and CCJT employ different transmission policies on the access channels, they use the same MMSE design for the backhaul to moderate the overhead signal among the ENs. In both cases, the ENs decodes the backhaul signal individually. 
\end{remark}
\begin{remark}
	In the cooperative architecture, the ENs employ different precoding designs on the access links, e.g., ZF, MMSE, and Optimal design, which results in different achievable rates on the access links, $R_{\mathtt{coop},k}$, and different ENs' transmit powers, $\sum_{k=1}^K{\parallel\!\bs{v}_k\!\parallel^2}$, which eventually affects the backhaul information rate.   
\end{remark}

%

%
%
%
\section{Delivery time minimization in Distributed caching mode}\label{sec:Dist}
In this section, we propose a power allocation to minimize the delivery time in DCST mode. 
%
For an EN which has the whole requested files in its cache, e.g., EN $k\notin \mathcal{U}_C$, the delivery time for this EN to serve its user is $t_k = \frac{Q}{R_k}, \forall k \notin \mathcal{U}_C$, i.e., $K_C < k \leq K$, where $Q$ is the file size.

In order to serve a user $k \in \mathcal{U}_C$, the EN $k$ will receive the non-cached parts from the WAP while serving its user in the FD mode. Assuming that the FastForward FD transmission is employed by the ENs \cite{Bharadia14:FFC}, the delivery time for the user $k$, $\forall k\in \mathcal{U}_C$, is $t_k = \frac{Q}{R_k}$ subjected to a constraint that the EN's buffer is not empty. Because $\mu_{d_k}Q$ bits of the requested file is already available at the EN $k$'s cache, this condition reads $C_{\mathtt{dist},k} \tau + \mu_{d_k}Q \geq R_k \tau, \forall \tau \in [0, t_k]$. Consider all possible values of $\tau \in [0, t_k]$, this constraint becomes $C_{\mathtt{dist},k} \geq \bar{\mu}_kR_k, \forall k\in \mathcal{U}_C$, where $\bar{\mu}_k\triangleq 1 - \mu_{d_k}$ represents the volume of the non-cached parts of the requested file $f_{d_k}$. 

We would like to minimize the largest delivery time among the users. The optimization problem is formulated as follows:
\begin{align}\label{OP:Tdist}
	\underset{\{p_k\}_{k=1}^K, \{q_k\}_{k=1}^{K_C}}{\mathtt{minimize}} &~~ \max \Big(\frac{Q}{R_1}, \dots, \frac{Q}{R_K} \Big), \\
	\mathtt{s.t.} 
	~ C_{\mathtt{dist},k} \geq& \bar{\mu}_k R_k, \forall k\in \mathcal{U}_C \subeqn \label{eq:Tdist c1}\\
	~ \sum_{k \in \mathcal{U}_C}\!\! \parallel\! \tilde{\bs{w}}_k&\!\parallel^2 q_k \le P_{BS};
	p_k \leq P_{EN}, \forall k, \subeqn \label{eq:Tdist c2}
\end{align}
where the first constraint is to guarantee the EN's cache is not empty, $P_{BS}$ and $P_{EN}$ are the maximum transmit power at the WAP and the ENs, respectively. Although the objective function of problem \eqref{OP:Tdist} can be transformed into the max-min rate problem, the key challenge lies in the non-convexity of constraint \eqref{eq:Tdist c1}.

For ease of presentation, let $\bs{p} = [p_1, \dots, p_K, 1]^T$ and $\bs{q} = [q_1, \dots, q_{K_C}, 1]^T$ denote the compound power variables. In addition, we define following parameters:
\begin{align*}
	A_{1k} =& [|\bs{g}^H_k\tilde{\bs{g}}_1|^2, \dots, |\bs{g}^H_k\tilde{\bs{g}}_{K_C}|^2, \sigma^2]\\
	A_{2k} =& \\ 
	[|\bs{g}^H_k\tilde{\bs{g}}_1&|^2, \dots, |\bs{g}^H_k\tilde{\bs{g}}_{k-1}|^2, 0, |\bs{g}^H_k\tilde{\bs{g}}_{k+1}|^2, \dots, |\bs{g}^H_k\tilde{\bs{g}}_{K_C}|^2, \sigma^2]\\
	B_{1k} =& [|h_{k1}|^2, \dots, |h_{kN}|^2, \sigma^2]\\
	B_{2k} =& [|h_{k1}|^2, \dots, |h_{k(k-1)}|^2, 0, |h_{k(k+1)}|^2, \dots,  |h_{kN}|^2, \sigma^2]\\
	D_k =& [|f_{k1}|^2, \dots, |f_{k(k-1)}|^2, \eta, |f_{k(k+1)}|^2,\dots, |f_{kN}|^2, 0] \\
	\bs{\lambda} =& [{\parallel \tilde{\bs{g}}_1\parallel}^2, \dots, {\parallel \tilde{\bs{g}}_{K_C}\parallel}^2, 0].
\end{align*}
From \eqref{eq:Ck dis} and \eqref{eq:Rk dis}, we can write then backhaul and access information rate as follows:
\begin{align}
	C_{\mathtt{dist},k} =& W\log_2\Big(\frac{D_k \bs{p} + A_{1k}\bs{q}}{D_k\bs{p} + A_{2k}\bs{q}} \Big)  = \label{eq:Ck dis1}\\
	 W\log_2&(D_k \bs{p}\! +\! A_{1k}\bs{q})\! -\! W\log_2(D_k \bs{p}\! +\! A_{2k}\bs{q}),\forall k  \in \mathcal{U}_C \notag \\
	R_{\mathtt{dist},k}\! =& W\log_2\Big(\frac{B_{1k}\bs{p}}{B_{2k}\bs{p}} \Big) \notag\\
	=& W\log_2(B_{1k}\bs{p}) - W\log_2(B_{2k}\bs{p}),\forall k. \label{eq:Rk dis1}
\end{align}

By introducing a positive variable $t$, and using \eqref{eq:Ck dis1} and \eqref{eq:Rk dis1}, the problem \eqref{OP:Tdist} is equivalent to the following problem:
%
%
%
\begin{align}
&	\underset{t, \bs{p}, \bs{q}}{\mathtt{minimize}}  ~~ t \label{OP:Tdist2}\\
	&\mathtt{s.t.}  ~~  \log(B_{1k} \bs{p}) \geq \frac{Q\log(2)}{Wt} + \log(B_{2k}\bs{p}), \forall k \subeqn \label{eq:T2 c1}\\
	&\qquad~ \log(D_k\bs{p} + A_{1k}\bs{q} ) + \bar{\mu}_k\log(B_{2k}\bs{p}) \notag \\
	&~~~~~~ \geq \bar{\mu}_k\log(B_{1k}\bs{p}) + \log(D_k\bs{p} + A_{2k}\bs{q}), \forall k\in \mathcal{U}_C \subeqn \label{eq:T2 c2}\\
	&~~~~~~ \bs{\lambda} \bs{q} \leq P_{BS}; ~p_k \leq P_{EN},\forall k, \subeqn \label{eq:T2 c3}
\end{align}
where the new constraint \eqref{eq:T2 c1} results from $t \ge \frac{Q}{R_{\mathtt{dist},k}}, \forall k$.

It is observed that problem \eqref{OP:Tdist2} is non-convex since the first two constraints are non-affine. To overcome this difficulty, we will represent these constraints in a convex expression via arbitrary intermediate variables $\{x_k, z_k\}_{k=1}^{K_C}$, $\{y_k\}_{k=1}^K$, and reformulate problem \eqref{OP:Tdist2} as follows:
\begin{align}
&\underset{t, \bs{p}, \bs{q}, \{y_k\}_{k=1}^K, \{x_k, z_k\}_{k=1}^{K_C}}{\mathtt{minimize}}  ~~ t \label{OP:Tdist3}\\
&\qquad\mathtt{s.t.} 
 ~~  \log(B_{1k} \bs{p}) \geq \frac{Q\log(2)}{W t} + y_k, \forall k \subeqn \label{eq:T3 c1}\\
&\qquad\qquad~ \log(D_k\bs{p} + A_{1k}\bs{q}) + \bar{\mu}_k\log(B_{2k}\bs{p}) \notag \\
&\qquad\qquad~~ \geq \bar{\mu}_k x_k + z_k,~1\leq k \leq K_C\subeqn \label{eq:T3 c2}\\
&\qquad\qquad~ B_{1k}\bs{p} \leq e^{x_k}, ~1\leq k \leq K_C \subeqn \label{eq:T3 c3}\\
&\qquad\qquad~ B_{2k}\bs{p} \leq e^{y_k}, \forall k \subeqn \label{eq:T3 c4}\\
&\qquad\qquad~ D_k\bs{p}+A_{2k}\bs{q} \leq e^{z_k},~ 1\leq k \leq K_C \subeqn \label{eq:T3 c5}\\
&\qquad\qquad~ \bs{\lambda} \bs{q} \leq P_{BS}; ~p_k \leq P_{EN},\forall k. \subeqn \label{eq:T3 c6}
\end{align}
Although constraints \eqref{eq:T3 c1} and \eqref{eq:T3 c2} are now convex, solving problem \eqref{OP:Tdist3} is still challenging since constraints \eqref{eq:T3 c3} - \eqref{eq:T3 c5} are unbounded. Fortunately, because the function $e^x$ is convex, we can employ the inner approximation method, which replaces constraints \eqref{eq:T3 c3} - \eqref{eq:T3 c5} by using the first-order approximation of the exponential function, i.e., $e^x \simeq e^{x_0}(x - x_0 + 1)$, where $x_0$ is any accessible point. The approximated problem is formulated, for a given set of accessible points $\bs{x}_0 \triangleq \{x_{0k}\}_{k=1}^{K_C}, \bs{y}_0 \triangleq \{y_{0k}\}_{k=1}^K, \bs{z}_0 \triangleq \{z_{0k}\}_{k=1}^{K_C}$, as follows:
\begin{align}
\bs{Q}_1(&\bs{x}_0,\bs{y}_0,\bs{z}_0): \underset{t, \bs{p}, \bs{q}, \{x_k,z_k\}_{k=1}^{K_C},\{y_k\}_{k=1}^K}{\mathtt{minimize}}  ~~ t \label{OP:Tdist4}\\
\mathtt{s.t.} 
& ~~  \eqref{eq:T3 c1},~\eqref{eq:T3 c2},~\eqref{eq:T3 c6}\notag\\
&~~ B_{1k}\bs{p} \leq e^{x_{0k}}(x_k - x_{0k} + 1), 1\leq k \leq K_C \subeqn \label{eq:T4 c1}\\
&~~ B_{2k}\bs{p} \leq e^{y_{0k}}(y_k - y_{0k} + 1), \forall k \subeqn \label{eq:T4 c2}\\
&~~ A_{2k}\bs{q}\! +\! D_k\bs{p} \leq e^{z_{0k}}(z_k\! -\! z_{0k}\! +\! 1), 1\leq k \leq K_C. \subeqn \label{eq:T4 c3}
\end{align}
%

It is straightforward to verify that, for a given set of $\bs{x}_0,\bs{y}_0,\bs{z}_0$,  problem \eqref{OP:Tdist4} is convex since the objective function and the constraints are convex. Thus, it can be solved in an efficient manner by standard solvers, e.g., CVX. Since $e^{x_0}(x - x_0 + 1) \leq e^{x}, \forall x_0$, the approximated problem \eqref{OP:Tdist4} always gives a suboptimal solution of the original problem \eqref{OP:Tdist3}.

We note that the optimal solution of problem \eqref{OP:Tdist4} is largely determined by the parameters $\{x_{0k},  z_{0k}\}_{k=1}^{K_C}, \{y_{0k}\}_{k=1}^K$. Therefore, it is important to choose proper values $\{x_{0k}, y_{0k}, z_{0k}\}$ such that the solution of \eqref{OP:Tdist4} approaches quickly the optimal solution of \eqref{OP:Tdist3}. As such, we propose an iterative optimization algorithm to improve the performance of problem \eqref{OP:Tdist4}. The premise behind the proposed algorithm is to better select the parameters $\{x_{0k},  z_{0k}\}_{k=1}^{K_C}, \{y_{0k}\}_{k=1}^K$ through iterations. The details of the proposed algorithm are presented in Table~\ref{tab:1}.

The convergence of the proposed iterative algorithm is guaranteed in the proposition below.
\begin{proposition}\label{prop 1}
The objective function of problem $\bs{Q}_1(\bs{x}_0,\bs{y}_0, \bs{z}_0)$ in (\ref{OP:Tdist4}) solved by the iterative algorithm in Table~\ref{tab:1} decreases by iterations.
\end{proposition}
\begin{IEEEproof}
	See Appendix~\ref{app:1}.
\end{IEEEproof}
Although Proposition~\ref{prop 1} does not guarantee the optimality of the approximated problem, it provides justification for the proposed iterative optimization algorithm.

\begin{table}
	\centering
	\caption{\textsc{Iterative Algorithm to solve \eqref{OP:Tdist3}}}\label{tab:1}
	\begin{tabular}{l l}
		\hline
		\vspace{-0.0in} & \\
		1. & Initialize $\bs{x}_0, \bs{y}_0, \bs{z}_0$, $\epsilon$, $t_{\rm old}$ and $\mathtt{error}$. \\
		2. & While $\mathtt{error} > \epsilon$ do \\
		&  2.1. Solve $\bs{Q}(\bs{x}_0,\bs{y}_0,\bs{z}_0)$ in \eqref{OP:Tdist4} to obtain the optimal \\
		&\ \ \ \ values $t_\star, \bs{p}_\star, \bs{q}_\star, \bs{x}_\star, \bs{y}_\star, \bs{z}_\star$\\
		& 2.3. Compute $\mathtt{error} = |t_\star - t_{\rm old}|$\\
		& 2.4. Update $t_{\rm old} = t_\star, \bs{x}_0 = \bs{x}_\star, \bs{y}_0 = \bs{y}_\star, \bs{z}_0 = \bs{z}_\star$\\
		\hline
	\end{tabular} \vspace{-0.1in}
\end{table}
\section{Delivery time minimization in Cooperative caching mode}\label{sec:coop}
In this section, we minimize the delivery time under the CCJT mode. Intuitively, the cooperative caching mode not only reduces inter-user interference on the access links, but also improves the self-interference cancellation at the ENs since the ENs' transmit signals are shared among the ENs. 

{We consider three precoding designs for the access links: ZF, MMSE and optimal design which jointly optimizes the direction and magnitude of the precoding vectors. We note that the WAP employs the same backhaul precoding design as in Section~\ref{sec:Dist}}.
\subsection{Delivery time minimization under ZF design}\label{sec:ZF}
The precoding vector under the ZF design is given as $\bs{v}_k = \sqrt{p_k} \breve{\bs{h}}_k$, where $p_k$ is the power factor allocated for user $k$ and  $\bs{\breve{h}}_k$ is the ZF beamforming vector, which is the $k$-th column of the ZF precoding matrix $\bs{H}^H (\bs{H}\bs{H}^H)^{-1}$. In this design, the inter-user interference (on the access links) is fully cancelled, i.e., $|\bs{h}^H_k\breve{\bs{h}}_i|  = \delta_{ki}, \forall k,i$. From \eqref{eq:Ck dis} and \eqref{eq:Rk dis} we have the backhaul and access rates under the ZF design as follows:
\begin{align*}
C^{ZF}_{\mathtt{coop},k}\!\! &=\! \bts W\!
\log_2\!\!\Big(\!1\! +\! \frac{|\bs{g}_k^H\tilde{\bs{w}}_k|^2q_k}{\underset{k\neq l \in \mathcal{U}_C}{\sum}\! |\bs{g}_k^H\! \tilde{\bs{w}}_l|^2q_l\!  +\! \eta \underset{i=1}{\overset{K}{\sum}} \parallel\!\breve{\bs{h}}_i\!\parallel^2p_i\! +\! \sigma^2} \!\Big), \forall k \in \mathcal{U}_C \ets  \\
R^{ZF}_{\mathtt{coop},k} &= W\log_2\Big(1 + \frac{p_k}{\sigma^2}\Big), \forall k. 
\end{align*}
The minimization problem of the largest delivery time under the ZF design is stated as follows:
\begin{align}\label{OP:Tzf}
\underset{\{p_k\}_{k=1}^K, \{q_k\}_{k=1}^{K_C}}{\mathtt{minimize}} &~~ \bts\max\ets (\frac{Q}{R^{ZF}_{\mathtt{coop},1}}, \dots, \frac{Q}{R^{ZF}_{\mathtt{coop},K}} ), \\
\mathtt{s.t.} 
&~~ \bts C^{ZF}_{\mathtt{coop},k} \geq \bar{\mu}_kR^{ZF}_{\mathtt{coop},k}, \forall k \in \mathcal{U}_C \ets\subeqn \label{eq:Tzf c1}\\
&~~ \bts{\sum}_{k\in \mathcal{U}_C} {\parallel \tilde{\bs{w}}_k\parallel}^2 q_k \le P_{BS}\ets \subeqn \label{eq:Tzf c2}\\
&~~ \bts{\sum}_{k=1}^K \parallel\!\breve{\bs{h}}_i\!\parallel^2p_k \le K P_{EN},\ets\subeqn \label{eq:Tzf c3}
\end{align}
where the constraint \eqref{eq:Tzf c3} benefits from power allocation among the ENs due to the ENs' joint transmission.

Denote $t = \max \Big(\frac{Q}{R^{ZF}_{\mathtt{coop},1}}, \dots, \frac{Q}{R^{ZF}_{\mathtt{coop},K}} \Big)$ as a new variable. Then problem \eqref{OP:Tzf} is equivalent to the following problem:
\begin{align}
&\underset{t, \{p_k\}_{k=1}^K, \{q_k\}_{k=1}^{K_C}}{\mathtt{minimize}}  ~~ t \label{OP:T zf}\\
\mathtt{s.t.} 
& ~~ \bts\log(1 + \frac{p_k}{\sigma^2}) \geq \frac{Q\log(2)}{W t}, \forall k \ets \subeqn \label{eq:T zf c1}\\
&~~ \bts\log\Big(1 + \frac{|\bs{g}^H_k\tilde{\bs{w}}_k|^2 q_k}{{\sum}_{k\neq l\in \mathcal{U}_C} |\bs{g}_k^H \tilde{\bs{w}}_l|^2q_l  + \eta \underset{i=1}{\overset{K}{\sum}} {\parallel\! \breve{\bs{h}}_i\!\parallel}^2 p_i + \sigma^2}\Big)\ets  \notag\\
&~~\bts \geq \bar{\mu}_k\log(1 + \frac{p_k}{\sigma^2}), 1 \le k \leq K_C \ets \subeqn\label{eq:T zf c2}\\
&~~ \eqref{eq:Tzf c2}, \eqref{eq:Tzf c3}. \notag
\end{align}

For ease of presentation, let us define parameters $A_{1k}, A_{2k}, \bs{\lambda}$ as in Sec.~\ref{sec:Dist}, and 
$
\bs{\alpha} \triangleq [{\parallel\!\breve{\bs{h}}_1\!\parallel}^2, \dots, {\parallel\!\breve{\bs{h}}_K\!\parallel}^2].
$

Furthermore, we use the compound notation for the powers $\bs{p} = [p_1, \dots, p_K]^T$ and $\bs{q} = [q_1, \dots, q_{K_C}, 1]^T$. 

Then the problem \eqref{OP:T zf}  can be reformulated as follows:
\begin{align}
\underset{t, \bs{p},\bs{q}}{\mathtt{minimize}} & ~~ t \label{OP:Tzf1}\\
\mathtt{s.t.} & ~~ \log(1 + \frac{p_k}{\sigma^2}) \geq \frac{Q\log(2)}{t W}, \forall k \subeqn \label{eq:Tzf1 c1}\\
&~~ \log(A_{1k}\bs{q} + \eta\bs{\alpha}\bs{p}) \geq \bar{\mu}_k\log(1+ \frac{p_k}{\sigma^2}) \notag\\
&~~~~     + \log(\eta\bs{\alpha}\bs{p} + A_{2k}\bs{q}), 1\leq k \leq K_C\subeqn \label{eq:Tzf1 c2}\\
&~~ \bs{\alpha}\bs{p} \leq KP_{EN};~ \bs{\gamma}\bs{q} \leq P_{BS}, \subeqn \label{eq:Tzf1 c3}
\end{align}

It is observed that problem \eqref{OP:Tzf1} is non-convex since the first two constraints are non-affine. By introducing arbitrary variables $\{x_k,y_k\}_{k=1}^{K_C}$, we can reformulate problem \eqref{OP:Tzf1} as
\begin{align}
&\underset{t, \bs{p},\bs{q},\{x_k, y_k\}_{k=1}^{K_C}}{\mathtt{minimize}}  ~~ t \label{OP:Tzf2}\\
&~~\mathtt{s.t.} ~~ \eqref{eq:Tzf1 c1}, \eqref{eq:Tzf1 c3} \notag\\
&~~~~~~~~ \log(A_{1k}\bs{q} + \eta\bs{\alpha}\bs{p}) \geq \bar{\mu}_k x_k + y_k, 1\leq k\le K_C\subeqn \label{eq:Tzf2 c1}\\
&~~~~~~~~ 1 + \frac{p_k}{\sigma^2} \le e^ {x_k}, 1\leq k\le K_C\subeqn \label{eq:Tzf2 c2}\\
&~~~~~~~~ \eta \bs{\alpha}\bs{p} + A_{2k}\bs{q} \le e^ {y_k}, 1\leq k\le K_C. \subeqn \label{eq:Tzf2 c3}
\end{align}
It is evident that problem \eqref{OP:Tzf2} is non-convex since the two last constraints \eqref{eq:Tzf2 c2} and \eqref{eq:Tzf2 c3} are unbounded. Similarly to the previous section, we employ the linear-approximation of the exponential function to approximate these two constraints. Let's $x_{0k}, y_{0k}$ be any accessible points, the constraints \eqref{eq:Tzf2 c2} and \eqref{eq:Tzf2 c3} can be approximated as follows:
\begin{align}
 1 + \frac{p_k}{\sigma^2} &\le e^ {x_{0k}}(x_k - x_{0k} + 1), 1\leq k\le K_C\subeqn \label{eq:Tzf2 c2 app}\\
\eta \bs{\alpha}\bs{p} + A_{2k}\bs{q} &\le e^ {y_{0k}}(y_k - y_{0k} + 1), 1\leq k\le K_C. \subeqn \label{eq:Tzf2 c3 app}
\end{align}
Then the problem \eqref{OP:Tzf2} can be approximated as
\begin{align}
\bs{Q}_2(\bs{x}_0, \bs{y}_0):~ &\underset{t, \bs{p},\bs{q},\{x_k, y_k\}_{k=1}^{K_C}}{\mathtt{minimize}}  ~~ t \label{OP:Tzf3}\\
&\mathtt{s.t.}  ~~ \eqref{eq:Tzf1 c1}, \eqref{eq:Tzf1 c3}, \eqref{eq:Tzf2 c1}, \eqref{eq:Tzf2 c2 app}, \eqref{eq:Tzf2 c3 app}, \notag
\end{align}
where $\bs{x}_0 \triangleq \{x_{0k}\}_{k=1}^{K_C}, \bs{y}_0 \triangleq \{y_{0k}\}_{k=1}^{K_C}$.

For a known feasible set $\{x_{0k}, y_{0k}\}_{k=1}^{K_C}$, it is evident that problem \eqref{OP:Tzf3} is convex, since the objective function and the constraints are convex. Hence, standard methods can be used to solve this problem effectively. We note that the approximated problem \eqref{OP:Tzf3} gives  a suboptimal solution of problem \eqref{OP:Tzf2} because $e^{x_{0k}}(x_k - x_{0k} + 1) \leq e^{x_k}, \forall x_{0k}$.

\begin{table}[t]
	\centering
	\caption{\textsc{Iterative Algorithm to solve \eqref{OP:Tzf2}}}\label{tab:2}
	\begin{tabular}{l l}
		\hline
		\vspace{-0.0in} & \\
		1. & Initialize $\bs{x}_0 \triangleq \{x_{0k}\}_{k=1}^{K_C}, \bs{y}_0 \triangleq \{y_{0k}\}_{k=1}^{K_C}$, $\epsilon$, $t_{\rm old}$ \\
		& and $\mathtt{error}$. \\
		2. & While $\mathtt{error} > \epsilon$ do \\
		&  2.1. Solve $\bs{Q}_2(\bs{x}_0,\bs{y}_0)$ in \eqref{OP:Tzf3} to obtain the optimal \\
		&\ \ \ \ values $t_\star, \bs{p}_\star, \bs{q}_\star, \bs{x}_\star, \bs{y}_\star$\\
		& 2.3. Compute $\mathtt{error} = |t_\star - t_{\rm old}|$\\
		& 2.4. Update $t_{\rm old} = t_\star, \bs{x}_0 = \bs{x}_\star, \bs{y}_0 = \bs{y}_\star$.\\
		\hline
	\end{tabular} \vspace{-0.0in}
\end{table}
Since the optimal solution of problem \eqref{OP:Tzf3} is influenced by the parameters $\bs{x}_0, \bs{y}_0$. An iterative optimization algorithm is proposed in Tab.~\ref{tab:2} to improve the performance of the approximated problem \eqref{OP:Tzf3}. The convergence of the proposed iterative algorithm is given in the following proposition.

\begin{proposition}\label{prop 2}
	The objective function of problem $\bs{Q}_2(\bs{x}_0,\bs{y}_0)$ in \eqref{OP:Tzf3} solved by the iterative algorithm in Table~\ref{tab:2} decreases by iterations.
\end{proposition}
\begin{IEEEproof}
	See Appendix~\ref{app:2}.
\end{IEEEproof}
It is evident from Proposition~\ref{prop 2} that the proposed optimization algorithm closes the gap between the approximated problem and the original problem as the number of iterations increases. 

\subsection{Delivery time minimization under MMSE design}
The precoding vector under the MMSE design is given as $\bs{v}_k = 
\sqrt{p_k} \tilde{\mathbf{h}}_k$, where $\tilde{\mathbf{h}}_k$ is the $k$-th column of the MMSE precoding matrix $ \bs{H}^H(\bs{H}\bs{H}^H + \sigma^2 \bs{I})^{-1}$. Substituting $\bs{v}_k$ into \eqref{eq:Ck dis} and \eqref{eq:Rk dis}, we obtain:
\begin{align*}
	C^{MSE}_{\mathtt{coop},k} &=\! W\!  
	\begingroup\textstyle\log_2\!\Big(1\! +\! \frac{|\bs{g}^H_k\tilde{\bs{w}}_k|^2q_k}
	{\underset{k\neq l \in \mathcal{U}_C}{\sum} |\bs{g}_k^H \tilde{\bs{w}}_l|^2q_l\!  +\! \eta \underset{i=1}{\overset{K}{\sum}} \parallel\!\tilde{\bs{h}}_i\!\parallel^2p_i\! +\! \sigma^2}\!\Big)\endgroup, \forall k\in \mathcal{U}_C \\
	R^{MSE}_{\mathtt{coop},k} &= \begingroup\textstyle W\log_2\Big(1 + \frac{|\bs{h}^H_k\tilde{\bs{h}}_k|^2p_k}{\underset{i\neq k}{\sum} |\bs{h}_k^H \tilde{\bs{h}}_i|^2p_i + \sigma^2}\Big)\endgroup, \forall k. 
\end{align*}

The minimization problem of the largest delivery time under the MMSE design is stated as follows:
\begin{align}\label{OP:Tmse}
\underset{\{p_k\}_{k=1}^K, \{q_k\}_{k=1}^{K_C}}{\mathtt{minimize}} &~~ \max \Big(\frac{Q}{R^{MSE}_{\mathtt{coop},1}}, \dots, \frac{Q}{R^{MSE}_{\mathtt{coop},K}} \Big), \\
\mathtt{s.t.} 
&~~ C^{MSE}_{\mathtt{coop},k} \geq \bar{\mu}_kR^{MSE}_{\mathtt{coop},k}, \forall k \in \mathcal{U}_C \subeqn \label{eq:Tmse c1}\\
&~~ {\sum}_{k\in \mathcal{U}_C} {\parallel \tilde{\bs{w}}_k\parallel}^2 q_k \le P_{BS} \subeqn \label{eq:Tmse c2}\\
&~~ {\sum}_{k=1}^K {\parallel \tilde{\bs{h}}_k\parallel}^2 p_k \leq KP_{EN}.\subeqn \label{eq:Tmse c3}
\end{align}
By using $C^{MSE}_{\mathtt{coop},k}, R^{MSE}_{\mathtt{coop},k}$ and introducing a new variable $t = \max \Big(\frac{Q}{R^{MSE}_{\mathtt{coop},1}}, \dots, \frac{Q}{R^{MSE}_{\mathtt{coop},K}} \Big)$, we can reformulated problem \eqref{OP:Tmse} as follows:
\begin{align}
&\underset{t, \{p_k, q_l\}}{\mathtt{minimize}}  ~~ t \label{OP:Tmse 1}\\
\mathtt{s.t.} 
& ~~ \bts\log\Big(1 + \frac{|\bs{h}^H_k\tilde{\bs{h}}_k|^2p_k}{\sum_{i\neq k}|\bs{h}^H_k\tilde{\bs{h}}_i|^2p_i + \sigma^2}\Big) \geq \frac{Q\log(2)}{t W}, \forall k \subeqn \ets \label{eq:Tmmse 1 c1}\\
&~~ \bts\log\Big(1 + \frac{|\bs{g}^H_k\tilde{\bs{w}}_k|^2q_k}{\underset{l\ne k}{\sum}|\bs{g}^H_k\tilde{\bs{w}}_l|^2q_l +\eta \underset{i=1}{\overset{K}{\sum}} {\parallel\!\tilde{\bs{h}}_i\!\parallel}^2p_i + \sigma^2}\Big) \geq \ets\notag\\
&~~ \bts \bar{\mu}_k\log\Big(1 + \frac{|\bs{h}^H_k\tilde{\bs{h}}_k|^2p_k}{\sum_{i\neq k}|\bs{h}^H_k\tilde{\bs{h}}_i|^2p_i + \sigma^2}\Big), \forall k \in \mathcal{U}_C \ets \subeqn \label{eq:Tmmse 1 c2}\\
&~~	\eqref{eq:Tmse c2}, \eqref{eq:Tmse c3}. \notag
\end{align}
In the next step, lets define parameters $A_{1k}, A_{2k}, \bs{\lambda}$ as in Sec.~\ref{sec:ZF}, and following parameters:
\begin{align*}
E_{1k} =& [|\bs{h}^H_k\tilde{\bs{h}}_1|^2, \dots, |\bs{h}^H_k\tilde{\bs{h}}_K|^2, \sigma^2]\\
E_{2k} =& \\
 [|\bs{h}^H_k\tilde{\bs{h}}_1&|^2, \dots, |\bs{h}^H_k\tilde{\bs{h}}_{k-1}|^2, 0, |\bs{h}^H_k\tilde{\bs{h}}_{k+1}|^2, \dots,
|\bs{h}^H_k\tilde{\bs{h}}_K|^2, \sigma^2]\\
\bs{\beta} =& [{\parallel \tilde{\bs{h}}_1\parallel}^2, \dots, {\parallel \tilde{\bs{h}}_K\parallel}^2, 0].
\end{align*}
Then, the problem \ref{OP:Tmse 1} can be reformulated as follows:
\begin{align}
&\underset{t, \bs{q}, \bs{p}}{\mathtt{minimize}}  ~~ t \label{OP:Tmse2}\\
\mathtt{s.t.} 
 &~ \log(E_{1k}\bs{p}) \geq \frac{Q\log(2)}{W t} + \log(E_{2k}\bs{p}), \forall k \subeqn \label{eq:mmse1 c1}\\
&~ \log(A_{1k}\bs{q} + \eta \bs{\beta} \bs{p}) + \bar{\mu}_k\log(E_{2k}\bs{p})  \notag\\
&~\geq \bar{\mu}_k\log(E_{1k}\bs{p}) + \log(A_{2k}\bs{q} + \eta \bs{\beta}\bs{p}), \forall k \in \mathcal{U}_C \subeqn \label{eq:mmse1 c2}\\
&~	\bs{\lambda} \bs{q} \leq P_{BS};~\bs{\beta}\bs{p} \le KP_{EN},\subeqn \label{eq:mmse1 c3}
\end{align}
where $\bs{p} = [p_1, \dots, p_K, 1]^T$ and $\bs{q} = [q_1, \dots, q_{K_C}, 1]^T$.

We observe that problem \eqref{OP:Tmse2} is in a similar form as problem \eqref{OP:Tdist2}, except the last constraint on the EN's transmit power. Since this constraint is linear, hence convex, we can employ the same technique in Sec.~\ref{sec:Dist} to solve \eqref{OP:Tmse2}. Obviously, the convergence of the iterative optimization algorithm solving \eqref{OP:Tmse2} is guaranteed by Proposition~\ref{prop 1}.

{\subsection{Delivery time minimization under optimal precoding design}
In this subsection, we minimize the delivery time via general (and optimal) precoding design on the access links which jointly optimizes both direction and magnitude of the beamforming vectors $\bs{v}_k \in \mathbb{C}^{K\times 1}, \forall k$. The backhaul and access rate in this case are given as }
{
\begin{align*}
C^{Opt}_{\mathtt{coop},k} &=\! W\!  
\begingroup\textstyle\log_2\!\Big(1\! +\! \frac{|\bs{g}^H\tilde{\bs{w}}_k|^2q_k}
{\underset{k\neq l \in \mathcal{U}_C}{\sum} |\bs{g}_k^H \tilde{\bs{w}}_l|^2q_l\!  +\! \eta \underset{i=1}{\overset{K}{\sum}} \|\bs{v}_k \|^2\! +\! \sigma^2}\!\Big)\endgroup, \forall k\in \mathcal{U}_C \\
R^{Opt}_{\mathtt{coop},k} &= \begingroup\textstyle W\log_2\Big(1 + \frac{|\bs{h}^H\bs{v}_k|^2}{\underset{i\neq k}{\sum} |\bs{h}_k^H \bs{v}_i|^2 + \sigma^2}\Big)\endgroup, \forall k. 
\end{align*} 
The delivery time minimization problem under the optimal design is formulated as follows:
\begin{align}\label{OP:Topt}
\underset{\{\bs{v}_k\}_{k=1}^K, \{q_k\}_{k=1}^{K_C}}{\mathtt{minimize}} &~~ \max \Big(\frac{Q}{R^{Opt}_{\mathtt{coop},1}}, \dots, \frac{Q}{R^{Opt}_{\mathtt{coop},K}} \Big), \\
\mathtt{s.t.} 
&~~ C^{Opt}_{\mathtt{coop},k} \geq \bar{\mu}_kR^{Opt}_{\mathtt{coop},k}, \forall k \in \mathcal{U}_C \subeqn \label{eq:Topt c1}\\
&~~ {\sum}_{k\in \mathcal{U}_C} {\parallel \tilde{\bs{w}}_k\parallel}^2 q_k \le P_{BS} \subeqn \label{eq:Topt c2}\\
&~~ {\sum}_{k=1}^K \|\bs{v}_k\|^2 \leq KP_{EN}.\subeqn \label{eq:Topt c3}
\end{align}
The challenge in solving \eqref{OP:Topt} lies in the appearance of $\|\bs{v}_k\|^2$ in the denominator of both backhaul and access rates. To leverage this difficulty, we introduce new variables $\bs{V}_k \triangleq \bs{v}_k^H\bs{v}_k \in \mathbb{C}^{K\times K}$, which is symmetric and positive definite. It is straightforward to verify that $\|\bs{v}_k\|^2 = \Tr(\bs{V}_k)$ and $|\bs{h}_k^H \bs{v}_i|^2 = \Tr(\bs{H}_k \bs{V}_i)$, where $\bs{H}_k \triangleq \bs{h}^H_k\bs{h}_k$. Furthermore, by using a slack variable $t$ we can equivalently reformulate problem \eqref{OP:Topt} similarly to the previous subsection as
\begin{align}
&\underset{t, \{\bs{V}_k, q_l\}}{\mathtt{minimize}}  ~~ t \label{OP:Topt 1}\\
\mathtt{s.t.} 
& ~~ \bts\log\Big(1 + \frac{\Tr(\bs{H}_k\bs{V}_k)}{\sum_{i\neq k} \Tr(\bs{H}_k\bs{V}_i) + \sigma^2}\Big) \geq \frac{Q\log(2)}{t W}, \forall k \subeqn \ets \label{eq:Topt 1 c1}\\
&~~ \bts\log\Big(1 + \frac{|\bs{g}^H_k\tilde{\bs{w}}_k|^2q_k}{\underset{l\ne k} {\sum}|\bs{g}^H_k\tilde{\bs{w}}_l|^2q_l + \underset{i=1}{\overset{K}{\sum}} \Tr(\bs{V}_i) + \sigma^2}\Big) \geq \ets\notag\\
&~~ \bts \bar{\mu}_k\log\Big(1 + \frac{\Tr(\bs{H}_k\bs{V}_k)} {\sum_{i\neq k}\Tr(\bs{H}_k\bs{V}_i) + \sigma^2}\Big), \forall k \in \mathcal{U}_C \ets \subeqn \label{eq:Topt 1 c2}\\
&~~\sum_{k=1}^K \Tr(\bs{V}_k) \le KP_{EN};~\mathrm{rank}(\bs{V}_k) = 1, \forall k, \subeqn \label{eq:Topt 1 c3}\\
&~~	\eqref{eq:Topt c2}. \notag
\end{align}
By using similar notations $A_{1k}, A_{2k}, \bs{\lambda}$ as in Sec.~\ref{sec:ZF}, we can reformulate \eqref{OP:Topt 1} as
\begin{align}
&\underset{t, \{\bs{V}_k\}, \bs{q}}{\mathtt{minimize}}  ~~ t \label{OP:Topt 2}\\
\mathtt{s.t.} 
& ~~ \bts\log(\sum_{i=1}^K\Tr(\bs{H}_k\bs{V}_i) + \sigma^2) \geq \ets \subeqn \label{eq:Topt 2 c1} \\
&~~ \bts \qquad \frac{Q\log(2)}{t W} + \log(\sum_{i\neq k}\Tr(\bs{H}_k\bs{V}_i) + \sigma^2), \forall k \ets\notag\\
\bts \log\ets&\bts(A_{1k}\bs{q} + \eta \underset{i=1}{\overset{K}{\sum}}\Tr(\bs{V}_i)) + \bar{\mu} \log(\underset{i\neq k}{\sum}\Tr(\bs{H}_k\bs{V}_i) + \sigma^2) \geq \ets \notag\\
\bts\log\ets&\bts (A_{2k}\bs{q}\! +\! \eta \underset{i=1}{\overset{K}{\sum}}\Tr(\bs{V}_i))\! +\! \bar{\mu} \log(\underset{i=1}{\overset{K}{\sum}}\Tr(\bs{H}_k\bs{V}_i)\! +\! \sigma^2)\ets \subeqn \label{eq:Topt 2 c2}\\
&~~ \bs{\lambda}\bs{q} \le P_{BS};~ \sum_{k}\Tr(\bs{V}_k) \le KP_{EN} \subeqn \label{eq:Topt 2 c3}\\
&~~ \mathrm{rank}(\bs{V}_k) = 1, \forall k, \notag
\end{align}
where constraint \eqref{eq:Topt 2 c2} is applied only for $k \in \mathcal{U}_C$.
}

\begin{figure}[!b]
	\centering
	\subfigure[CCJT scheme]{\includegraphics[width=\columnwidth]{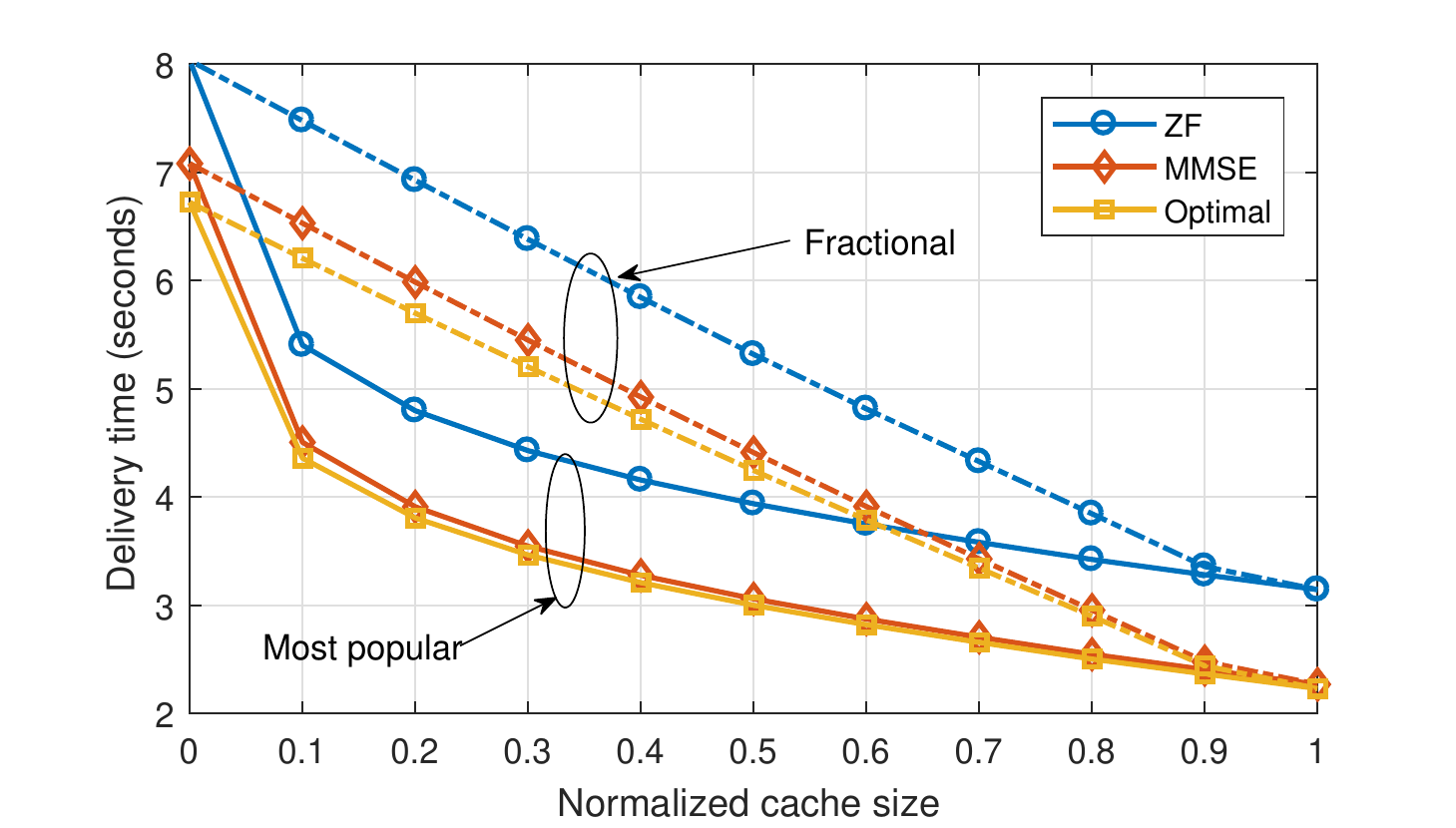}}
	\subfigure[DCST scheme]{\includegraphics[width=\columnwidth]{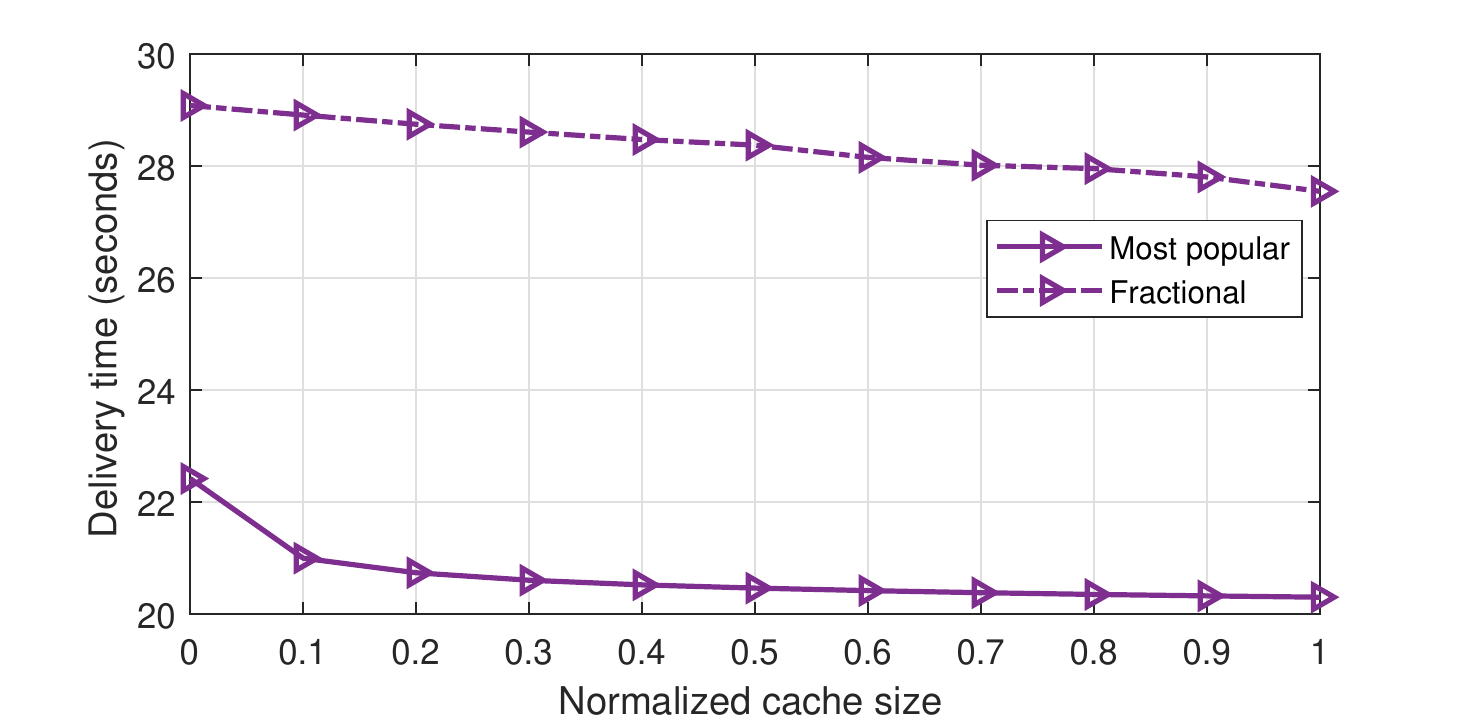}}
	\caption{Performance comparison of the proposed schemes under both  most popular and fractional caching policies. $P_{BS} = 3.16$W (5 dB) and $P_{EN} = 5$W.} \label{fig:ZipvsUni}
	\vspace{-0.2cm}
\end{figure}
{
Solving problem \eqref{OP:Topt 2} is difficult due to the non-convexity of \eqref{eq:Topt 2 c1}, \eqref{eq:Topt 2 c2} and the rank-one constraint. In order to deal with the latter, we employ the semidefinite relaxation (SDR) method \cite{Luo2010} which ignores the rank-one constraint when solving \eqref{OP:Topt 2}. SDR has been widely known as an efficient solution that achieves a close performance to the optimum \cite{Luo2010}\footnote{Since the SDR solution does not always guarantee the rank-one constraint, Gaussian randomization can be applied to improve the final performance. Details of Gaussian randomization technique are available in \cite{Luo2010}.}. To over the former, we observe that the trace function is linear and \eqref{eq:Topt 2 c1} and \eqref{eq:Topt 2 c2} are in similar form as constraints \eqref{eq:T2 c1} and \eqref{eq:T2 c2}, respectively. Therefore, we can can employ similar technique in Section~\ref{sec:Dist} to solve the SDR of \eqref{OP:Topt 2}, whose details are skipped to avoid redundancy.  
}
\section{Numerical results}\label{sec:results}
This section presents numerical results to demonstrate the effectiveness of our proposed optimization algorithms. The wireless channels are subject to Rayleigh fading. The pathloss on the backhaul is $G_1 = -60$dB. The pathloss on the access intended links is $G_2 = -50$dB. The pathloss on the inter-EN channels, e.g., $f_{kl}$, and the access interfering links, e.g., $h_{ki}, i\neq k$, are $G_E = -56$dB. Unless stated otherwise, the self-interference cancellation efficiency is equal to $\bar{\eta} = -70$dB \cite{Bharadia14:FDM}. Other parameters are as follows: $N = K = 4$, $\sigma^2 = -100$ dBm, $F = 100$ files, $Q = 100$Mb, and $W = 10$MHz. 
The simulation results are calculated based on 10000 random requests, equally distributed over 200 channel realizations. To achieve the best performance, we run the proposed iterative algorithms with 100 different initial values (see Table II and III for details) and select the best value. The user requests are assumed to follow the Zipf distribution with the skewness factor $\xi = 0.8$. {In the figures, we use $\rm{ZF}, \rm{MMSE}$ and $\rm{Optimal}$ to refer to ZF, MMSE and Optimal precoding designs, respectively}.

\subsection{Most popular caching versus fractional caching}

{
Fig.~\ref{fig:ZipvsUni} presents the delivery time performance of the proposed CCJT (a) and DCST (b) as a function of the normalized cache size, the ratio of the cache size divided by the library size, i.e., $\frac{M}{F}$. Both the most popular and fractional caching policies are presented. In the former, the most $M$ popular files are prefetched in the EN's cache, while in the later, a portion $\frac{M}{F}$ of every files are cached. In general, the most popular caching policy spends less time to serve the user requests than the fractional caching in both CCJT and DCST schemes. This is because the user requests follow a Zipf-based distribution, in which popular files are requested more frequently than the less popular ones. Since the most popular caching policy is more efficient than the fractional caching strategy, we only present the results for the most popular caching in the rest of the paper.
}

\subsection{Effectiveness of the proposed optimization algorithms and cooperative caching}
\begin{figure}
	\centering
	\subfigure[CCJT scheme]{\includegraphics[width=\columnwidth]{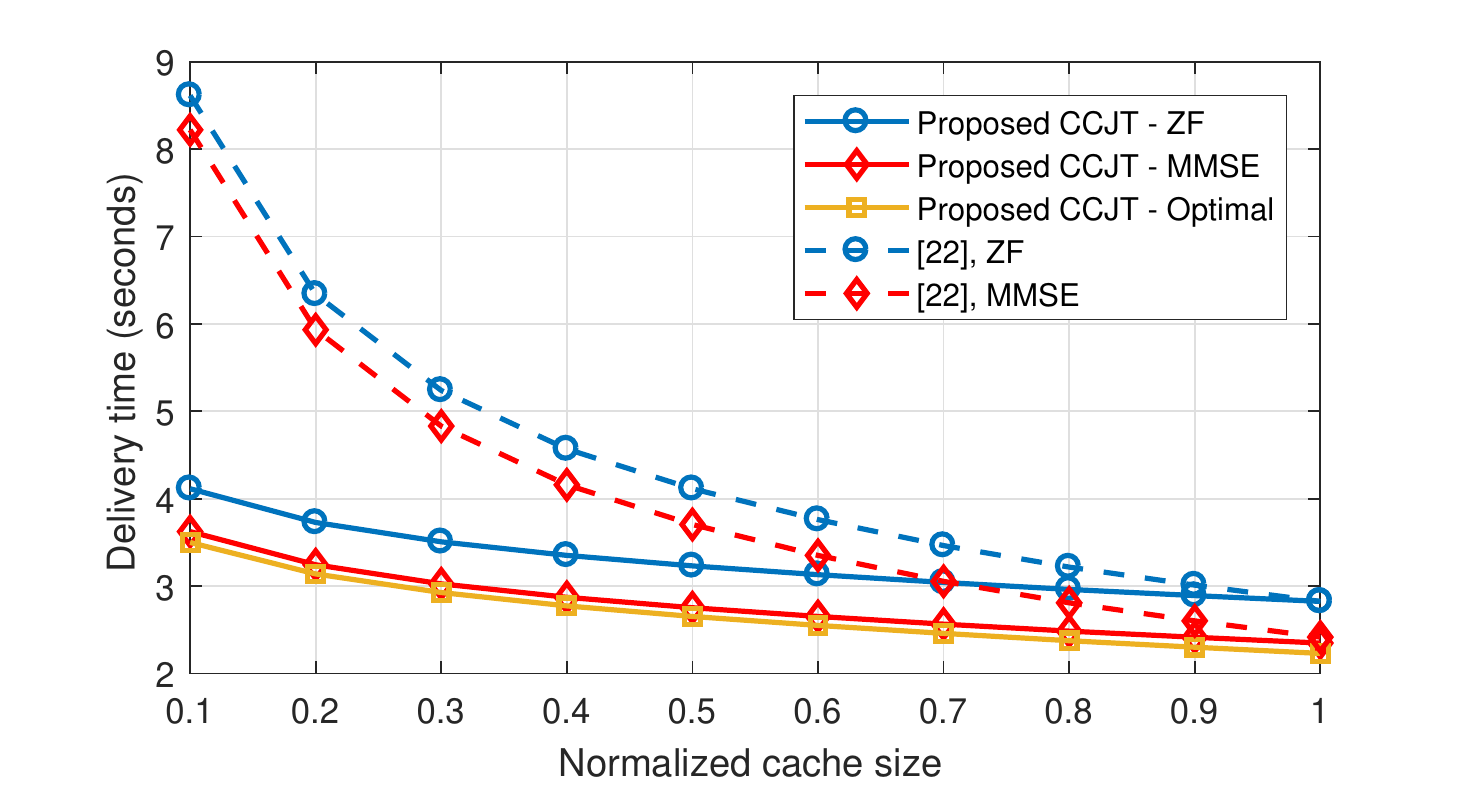}}
	\subfigure[DCST scheme]{\includegraphics[width=\columnwidth]{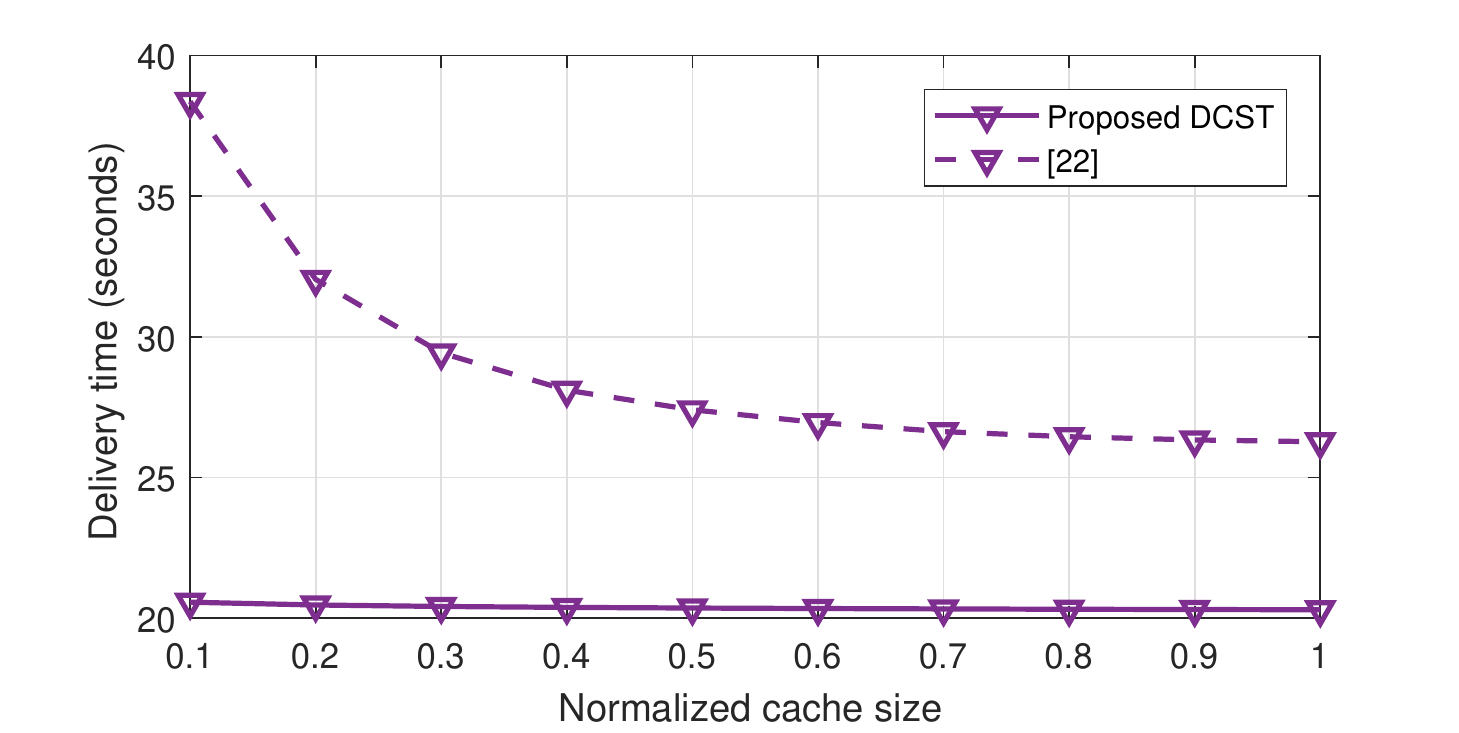}}
	\caption{The caching gain in FD-MEC systems v.s. the WAP's transmit power. EN's transmit power $P_{EN} = 5$W, the WAP's transmit pwoer $P_{BS} = 10$W.} \label{fig:OptimizationGain}
	\vspace{-0.2cm}
\end{figure}
We compare the proposed FD-MEC optimization algorithms with \cite{Marso17:CacheFD}, which proposes a FD-aided edge caching scheme with static transmit power. Although \cite{Marso17:CacheFD} considers only DCST, this method can be directly applied to CCJT under linear precoding designs without power control. 
In Fig.~\ref{fig:OptimizationGain}, we demonstrate the effectiveness of the proposed optimization algorithms in both CCJT (a) and DCST (b). It is noted that the reference \cite{Marso17:CacheFD} under linear precoding designs, i.e., ZF and MMSE, always transmit at the maximum power, equally divided for the ENs on the backhaul and for users on the access channels. A large gain is observed for the proposed optimization algorithms compared to the reference, especially in the small and medium cache size regimes. At large cache sizes, most of the requested files will be available in the ENs' cache, hence less traffic on the backhaul is required. In this case, the equal-power mode achieves a close performance as the proposed scheme. {Consider the precoding designs in CCJT,  the MMSE design performs considerably better than the ZF and achieves a close performance to the optimal precoding design. This is because MMSE and Optimal schemes perform power allocation more effectively than ZF, especially when the channel matrix is low rank. On average, the ZF design spends one second more than the tow others to serve the same demands. From a practical perspective, MMSE is preferred due to its low computation complexity compared with the Optimal scheme, as shown in Table~\ref{tab:4}}.

\begin{table}[h]
	\caption{{Average simulation time (in seconds) of three precoding designs, $K = 4$.}}
	\centering
	\begin{tabular}[Caption]{|c|c|c|}
		\hline
		ZF & MMSE & Optimal \\
		\hline
		0.0409 & 0.0509 & 0.1499\\
		\hline
	\end{tabular}
	\label{tab:4}
\end{table}

\begin{figure}
	\centering
	\includegraphics[width=\columnwidth]{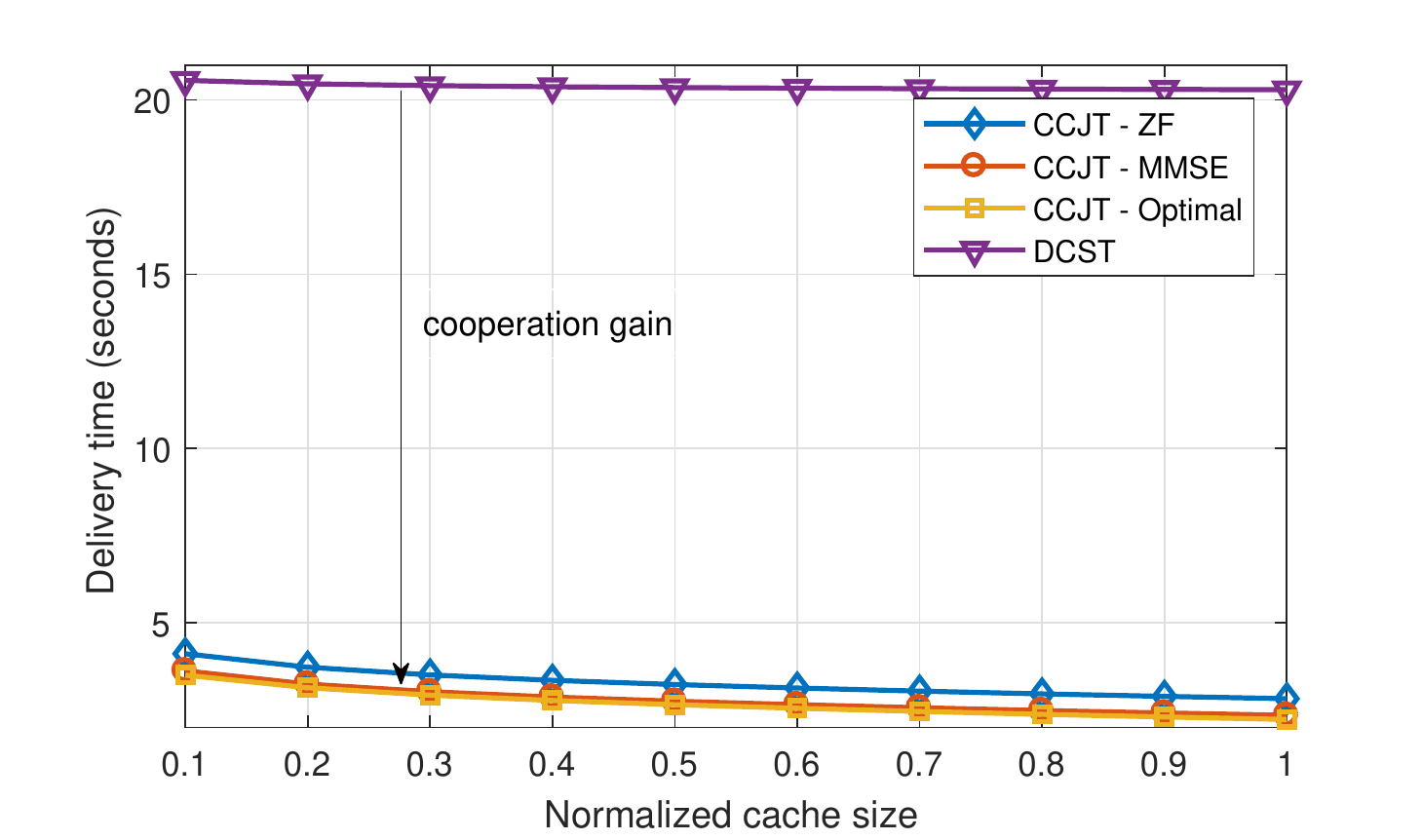}
	\caption{The caching gain in FD-MEC systems v.s. the WAP's transmit power. EN's transmit power $P_{EN} = 5$W, cache size $M = 0.4F$.} \label{fig:CooperGain}
	\vspace{-0.2cm}
\end{figure}

Fig.~\ref{fig:CooperGain} compares the delivery time of the CCJT with the DCST modes as a function of the normalized cache size. We recall that the DCST is fully decentralized and each EN operates independently. By allowing cooperative caching and joint transmission among the ENs, the delivery time dramatically drops for all cache sizes. In particular, the CCJT reduces the delivery time by about $85\%$ compared with DCST, which is mainly limited by both inter-EN and self interference. Obviously, this gain comes at the expense of extra physical connection and signal overheads among the ENs.

\begin{figure*}
	\normalsize
	\centering
	\subfigure[Absolute caching gain]{\includegraphics[width=\columnwidth]{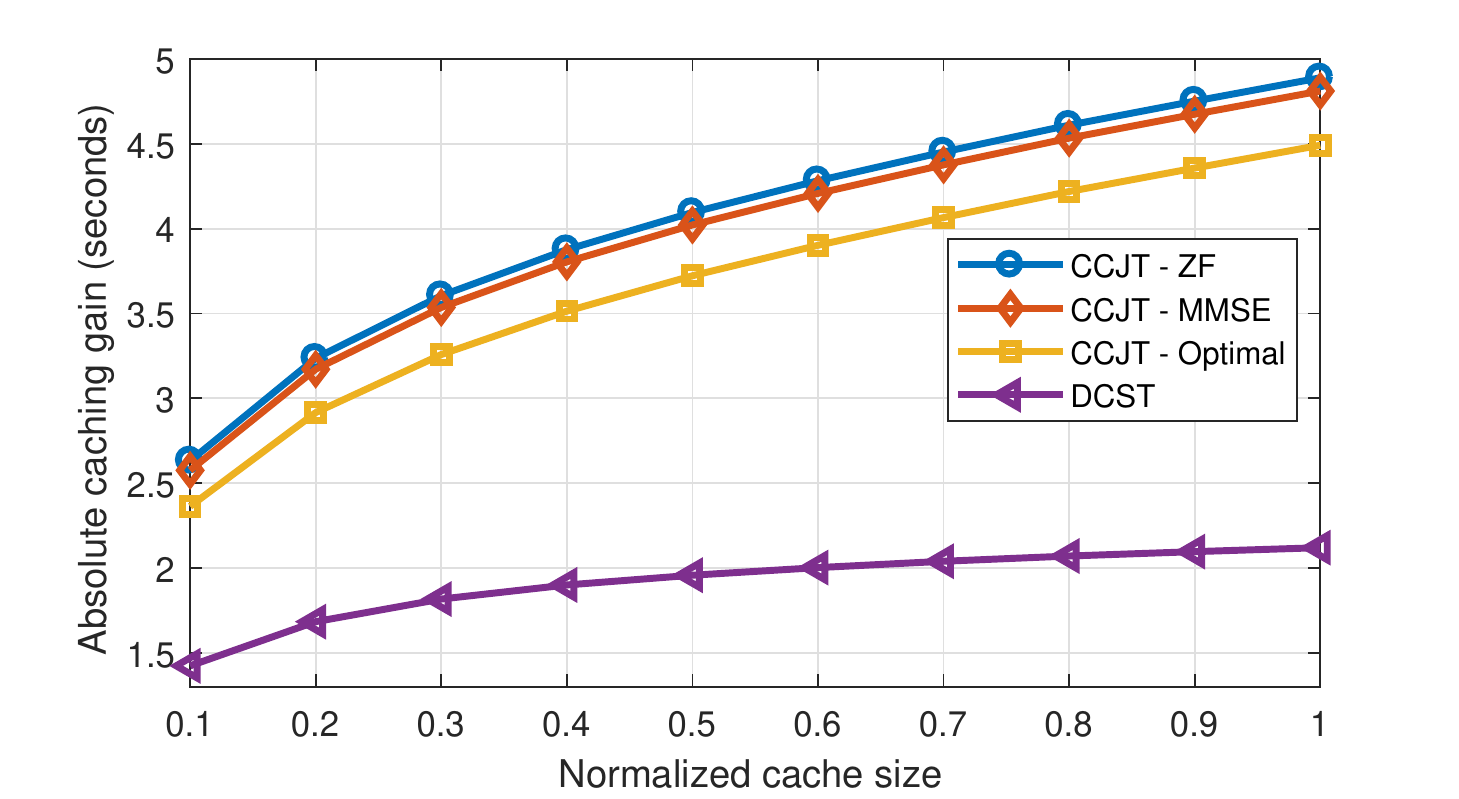}}
	\subfigure[Relative gain]{\includegraphics[width=\columnwidth]{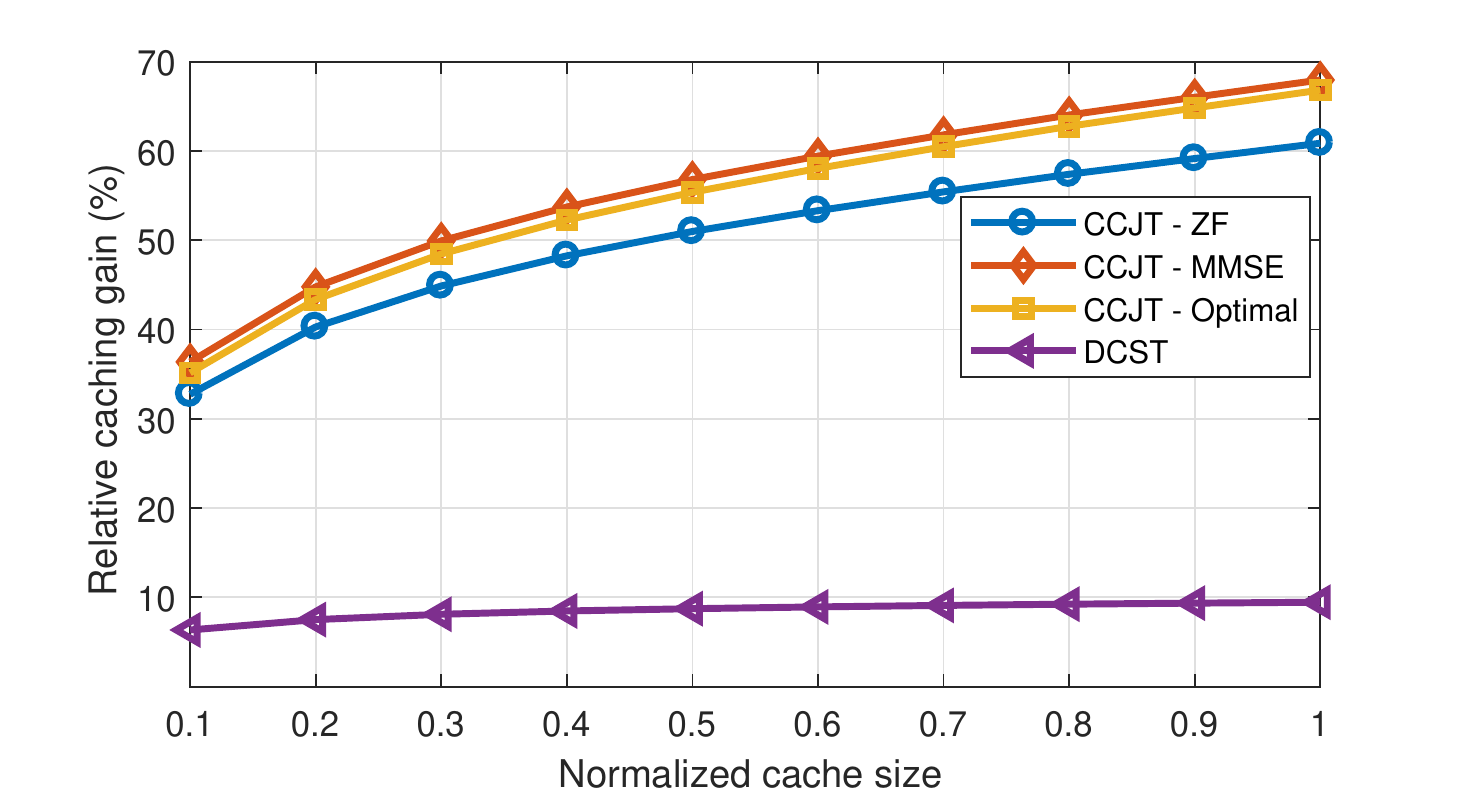}}
	\caption{{The caching gain in FD-MEC systems v.s. the normalized cache size $\frac{M}{F}$. $P_{BS} = 3.16$W (5 {dB}) and $P_{EN} = 5$W. Solid lines show the most popular caching policy. Dotted lines show the fractional caching policy.}} \label{fig:Gain_M}
	\vspace{-0.4cm}
\end{figure*}
\begin{figure*}
	\centering
	\subfigure[Absolute caching gain]{\includegraphics[width=\columnwidth]{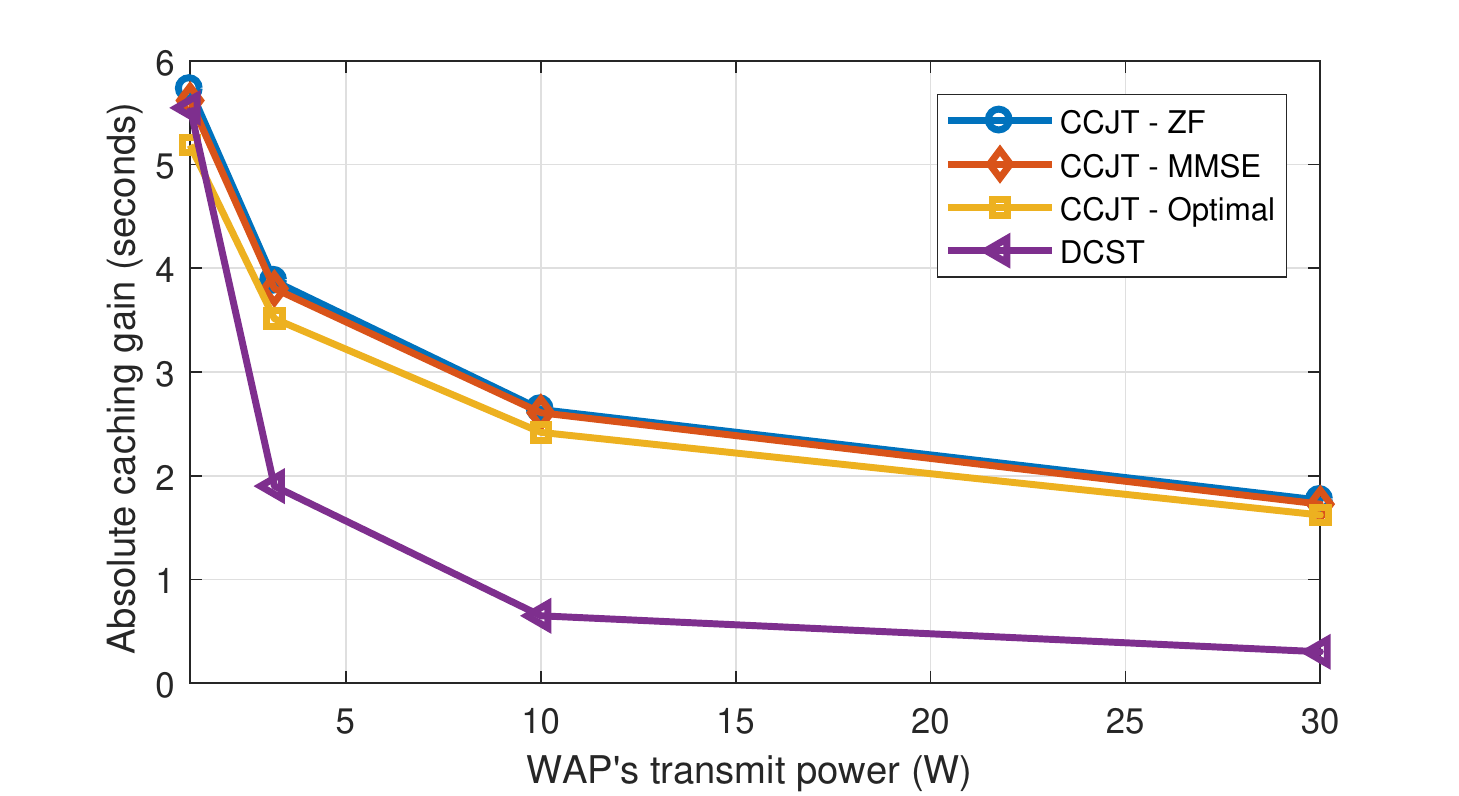}}
	\subfigure[Relative gain]{\includegraphics[width=\columnwidth]{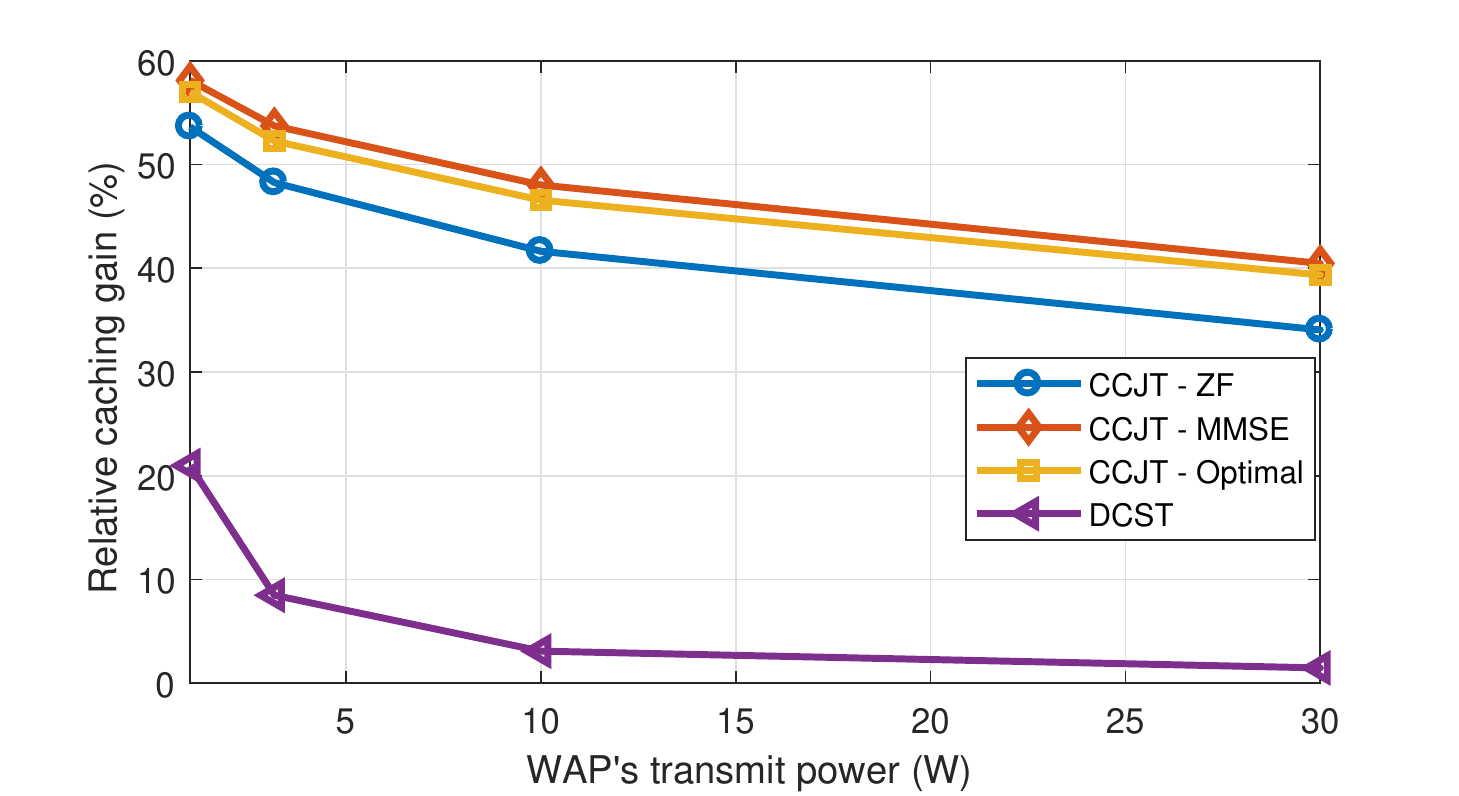}}
	\caption{The caching gain in FD-MEC systems v.s. the WAP's transmit power. EN's transmit power $P_{EN} = 5$W, cache size $M = 0.4F$.} \label{fig:Gain_Pbs}
	\vspace{-0.4cm}
\end{figure*}
\subsection{Role of caching in FD-MEC systems}
The effectiveness of caching in FD systems is demonstrated via a caching gain metric, which is computed as the delivery time reduction brought by the FD-MEC compared with the FD systems without caching capability at the ENs. In order to provide a complete observation, two types of caching gain are presented: \emph{Absolute caching gain} (ACG) and \emph{Relative caching gain} (RCG), which is calculated as follows:
\begin{align*}
	\mathrm{ACG} = t_{\rm no\ cache} - t_{\rm cache};~
	\mathrm{RCG} = 1 - \frac{t_{\rm cache}}{t_{\rm no\ cache}}, 
\end{align*}
where $t_{\rm cache}$ and $t_{\rm no\ cache}$ are the delivery time of the FD systems with and without caching at the ENs, respectively. 

Fig.~\ref{fig:Gain_M} presents the absolute caching gain (a) and relative caching gain (b) of the FD-MEC versus the normalized cache size. In general, caching in the cooperative mode CCJT is significantly more efficient than in the distributed architecture DCST. This expected outcome originates from two reasons. First, the shared cache among the ENs in CCJT facilitates the self-interference cancellation on the backhaul. Second, the joint transmission on the access links undoubtedly improves the access rates. At the cache size $M = 0.5F$, the CCJT (with all designs) achieves about 4 seconds reduction of the delivery time, twice as the DCST (Fig.~\ref{fig:Gain_M}a). The role of caching is shown more clearly via the relative caching gain in Fig.~\ref{fig:Gain_M}b: it reduces the delivery time by $55\%$ in the CCJT, compared with only $10\%$ in the DCST. We note that having the normalized cache size  equal $1$, i.e., $M = F$, does not result in $100\%$ relative caching gain since the total delivery time is lower bounded by the access channels. It is noted that although the ZF design achieves a larger absolute caching gain than MMSE and the Optimal, its relative gain is smaller. This implies that the ZF design is less efficient than the others.  

Fig.~\ref{fig:Gain_Pbs} shows the caching gains as a function of the WAP's transmit power, with $M = 0.4F$ and $P_{EN} = 5$W. A similar conclusion is drawn that CCJT is much more efficient than DCST. In addition, the influence of WAP's transmit power on the caching gain reduces as $P_{BS}$ becomes large. This is because at large WAP transmit powers, the delivery time is mainly determined by the access channel quality.

\begin{figure}
	\centering
	\subfigure[v.s. WPA's transmit power. $M = 0.4F$]{\includegraphics[width=\columnwidth]{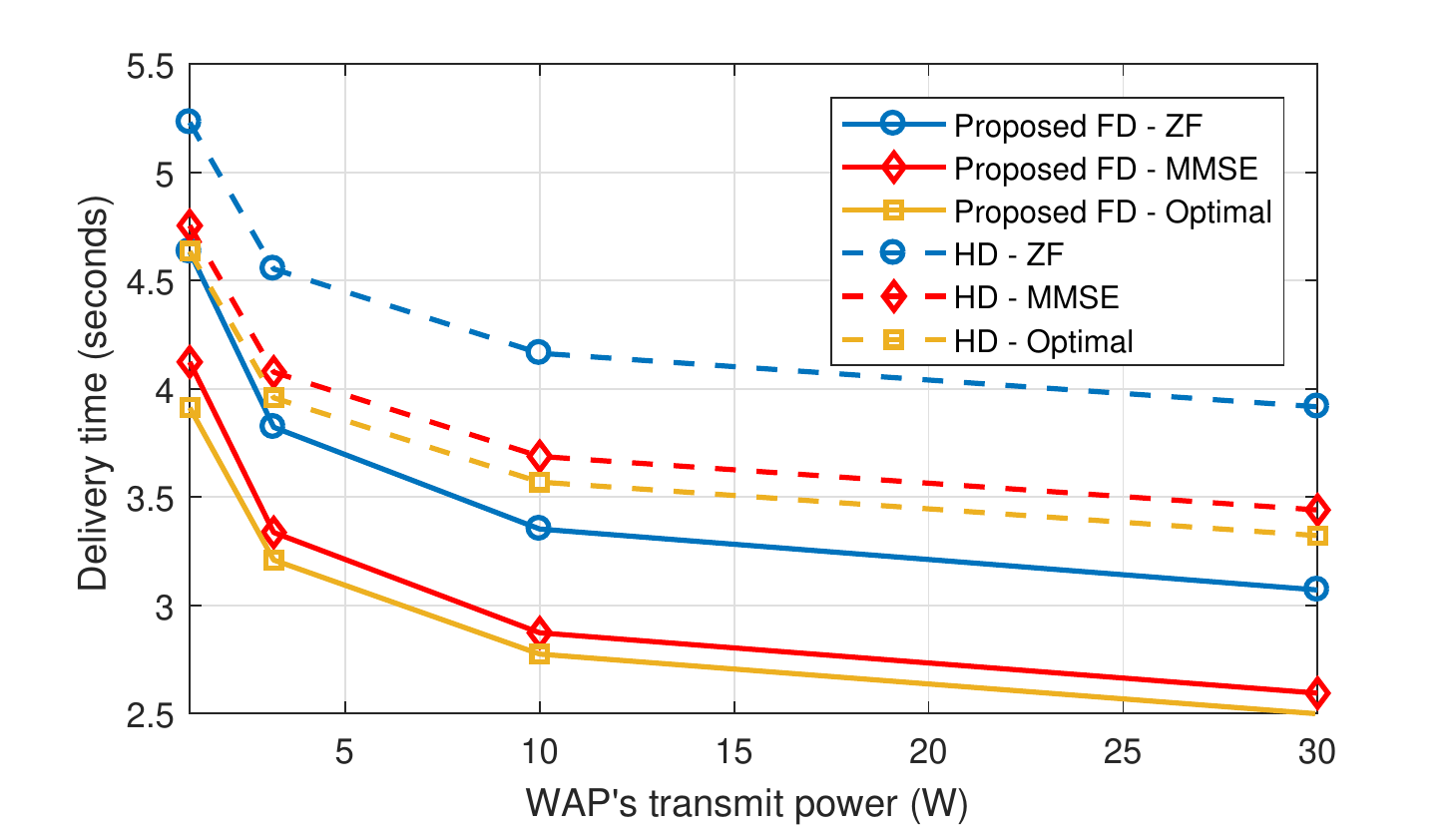}}
	\subfigure[v.s. normalized cache size. $P_{BS} = 10$W]{\includegraphics[width=\columnwidth]{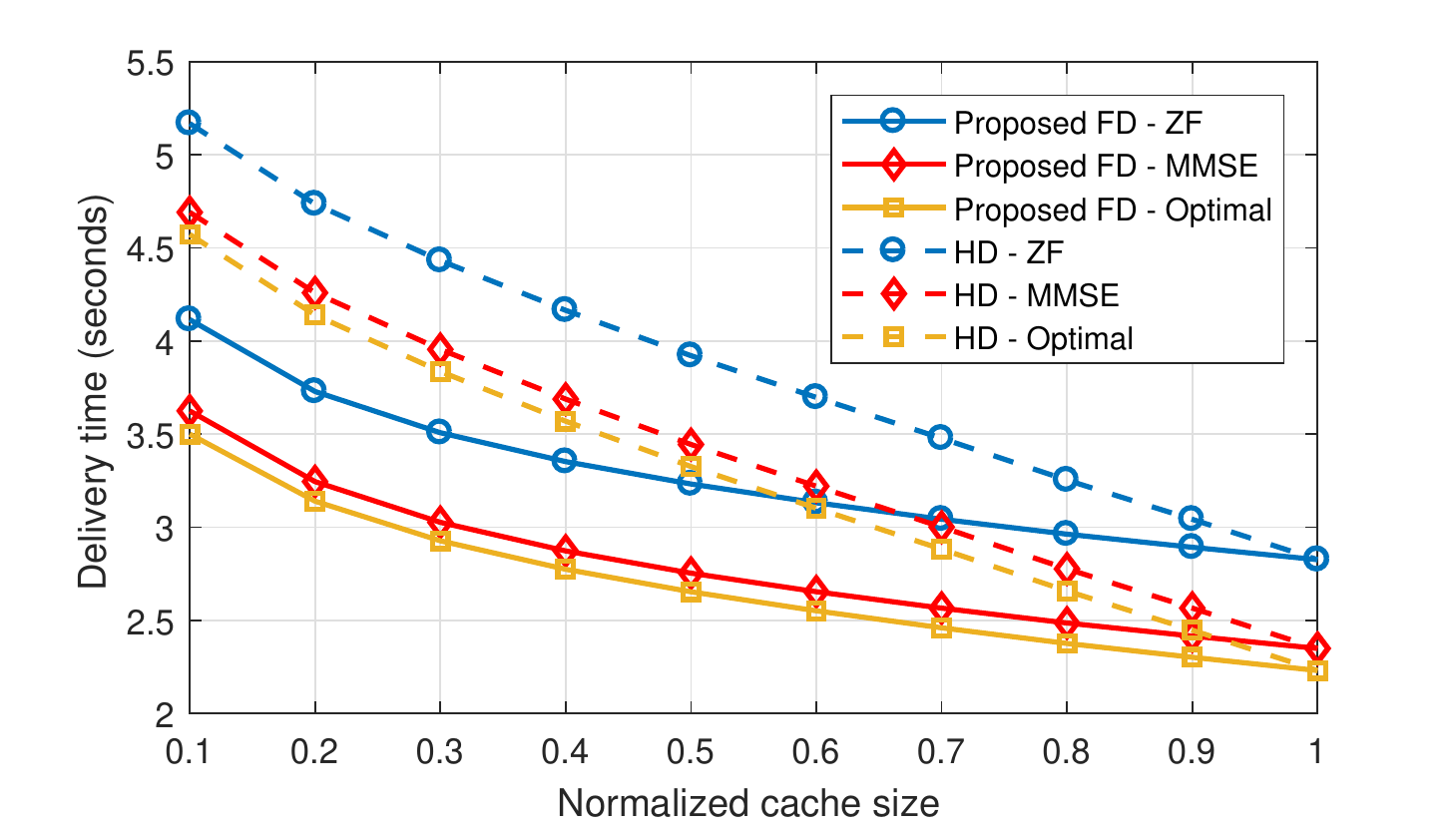}}
	\caption{Delivery time comparison between FD-MEC and HD system under the CCJT architecture. $P_{EN} = 5$W.} \label{fig:vsHD}
	\vspace{-0.2cm}
\end{figure}

\subsection{Comparison with half-duplex systems}
 In HD systems, the backhaul and access transmissions occur in two consecutive time slots. Therefore, the total delivery time in the HD mode is the summation of the delivery time on the backhaul link and on the access link. The delivery time of the HD mode is computed by the standard max-min design \cite{Vu18TWC}.
Fig.~\ref{fig:vsHD}a plots the delivery time as a function of the WAP's transmit power $P_{BS}$ under the CCJT mode, with $M = 0.4F$ and $P_{EN} = 5$W. It is observed that the FD-MEC system largely reduces the delivery time compared with the HD scheme for all precoding designs, i.e., ZF, MMSE and Optimal. At the WAP's transmit power equal to $5$W, a reduction of $25\%$ is obtained by the FD scheme for all precoding designs. An important observation is that large values of $P_{BS}$ will have less influence on the delivery time. In this case, increasing the WAP's transmit power does not lead to zero delivery time, since it is limited by the access link given a finite $P_{EN}$.

Fig.~\ref{fig:vsHD}b compares the delivery time of the FD-MEC with the HD system under the CCJT mode versus the normalized cache size, i.e., $\frac{M}{F}$. It is shown that the gain offered by the FD system over the HD is more significant in the small cache size ranges. The benefit of caching can be also interpreted as a means of trading memory for power: the delivery time with a large transmit power ($P_{BS} = 30$W, $M = 0.4F$ in Fig.~\ref{fig:vsHD}a) can also be achieved with a smaller transmit power and a larger cache size ($P_{BS} = 10$W, $M = 0.6F$ in Fig.~\ref{fig:vsHD}b). Increasing the cache size will diminish the advantage of the FD scheme over the HD. As such, it is highly probable that the requested file is already available at the EN's cache, thus there is less traffic on the backhaul. Note that having all the files cached does not result in zero delivery time due to the access link bottle neck. 

\begin{figure}
	\centering
	\includegraphics[width=\columnwidth]{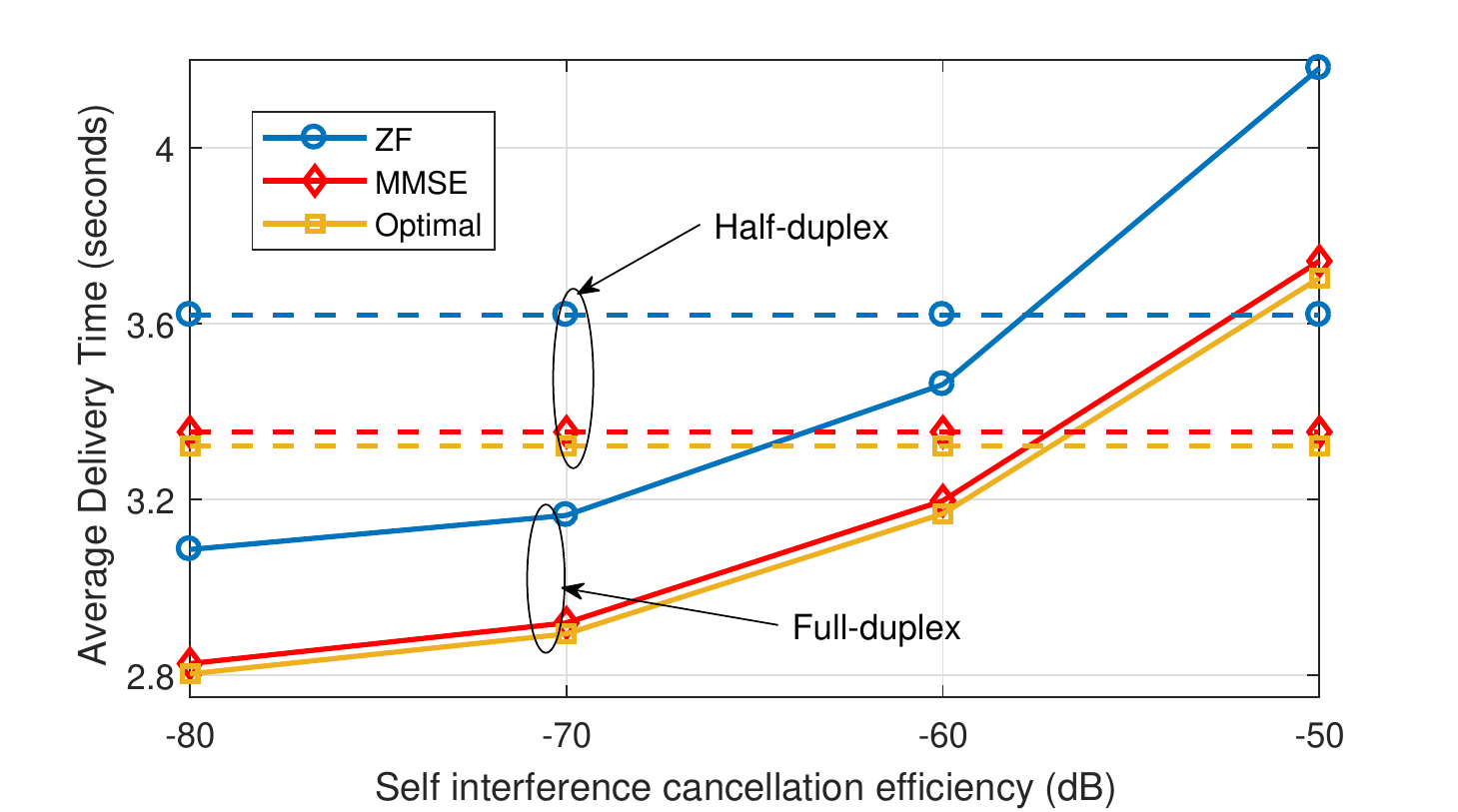}
	\caption{Average delivery time v.s. the self-interference cancellation efficiency $\bar{\eta}$. $M = 0.4F$, $P_{BS} = 10$W, and $P_{EN} = 5$W.} \label{fig:vseta}
	\vspace{-0.2cm}
\end{figure}

Fig.~\ref{fig:vseta} plots the delivery time versus the self-interference cancellation efficiency $\bar{\eta}$. Obviously, the delivery time of the HD system is independent from the cancellation efficiency since there is not self interference in this transmission mode. It is shown that the FD system outperforms the HD mode in the small values of $\bar{\eta}$. When the performance of the interference cancellation degrades, there is a crossing point between the FD and HD curves since the FD mode is limited by the residual interference. This result provides a guideline to determine the transmission mode when designing a cache-aided system. 

%
%
%
%
%
\section{Conclusions}\label{sec:conclusion}
In this paper, we have investigated the performance of full-duplex enabled mobile edge caching systems via the delivery time metric. The considered system is analysed under two  network architectures: distributed caching and cooperative caching. For each architecture, we proposed an optimal power control to minimize the system delivery time based on the linear precoding design. To overcome the non-convexity of the formulated problems, two iterative optimization algorithms have been proposed based on the inner approximation method, whose convergence is analytically guaranteed. {We have demonstrated that the cooperative caching perform largely better than the distributed scheme at the expense of full cooperation among the ENs. It has been also shown that the MMSE-based precoding design achieves the best trade-off between the performance and computation complexity. }

The considered schemes represent the two extremes of FD-MEC systems when collaboration among the ENs is available: i) the ENs operate in a complete decentralized manner, and ii) the ENs fully cooperate. Practical scenarios usually fall between these two modes. In this case, a cluster of ENs collaborate to serve their users, while the rest of the ENs operate independently. {One promising extension from this work is to optimize the caching policy at the ENs. This would require the derivation of the average delivery time over all fading channels.} 
%
%

\appendices
\section{Proof of Proposition~\ref{prop 1}}\label{app:1}
Denote $\big( t^{(i)}_\star, \bs{p}^{(i)}_\star, \bs{q}^{(i)}_\star, \bs{x}^{(i)}_\star, \bs{y}^{(i)}_\star, \bs{z}^{(i)}_\star  \big)$ as the optimal solution of $\bs{Q}_1(\bs{x}^{(i)}_0,\bs{y}^{(i)}_0, \bs{z}^{(i)}_0)$ at iteration $i$. We will show that if $x_{\star k}^{(i)} < x^{(t)}_{0k}, \forall k$, then by using $x^{(i+1)}_{0k} = x_{\star k}^{(i)}$ in the $(i+1)$-th iteration, we will have $t^{(i+1)}_\star  < t^{(i)}_\star $. Indeed, by choosing a relatively large initial value $\bs{x}^{(1)}_0,\bs{y}^{(1)}_0, \bs{z}^{(1)}_0$, we always have $x_{\star k}^{(1)} < x^{(1)}_{0k}, \forall k$. 

Denote $f(x;a) = e^a(x - a + 1)$ as the first order approximation of function $e^x$ at $a$. By using $\bs{x}^{(i)}_\star$ at the $(i+1)$-th iteration, we have $x^{(i+1)}_{0k} = x^{(i)}_{\star k}, \forall k$. Therefore, $f(x;x_{\star k}^{(i)})$ is used in the right-hand side of constraint \eqref{eq:T4 c1}. Consider a candidate $\bs{x}^{(i+1)} = \{x^{(i+1)}_1, \dots, x^{(i+1)}_{K_C}\}$, with $x^{(i+1)}_k = x_{\star k}^{(i)} - 1 + e^{x^{(i)}_{0k} - x_{\star k}^{(i)}}(x_{\star k}^{(i)} - x^{(i)}_{0k} + 1)$. It is evident that $x^{(i+1)}_k <  x_{\star k}^{(i)}$ and $f(x^{(i+1)}_k;x_{\star k}^{(i)}) = f(x_{\star k}^{(i)}; x^{(i)}_{0k}), \forall k \le K_C$. 

Because $x^{(i+1)}_k <  x_{\star k}^{(i)}, \forall k \le K_C$, the strictly inequality holds in constraint \eqref{eq:T3 c1}. Thus, there exits $t^{(i+1)} < t^{(i)}_{\star}$ which satisfies $\log(A_k\bs{p}) \geq \frac{Q\log(2)}{t^{(i+1)}} + x^{(i+1)}_k, \forall k$. Now consider a new candidate set $(t^{(i+1)}, \bs{p}^{(i)}_\star, \bs{q}^{(i)}_\star, \bs{x}^{(i+1)}, \bs{y}^{(i)}_\star, \bs{z}^{(i)}_\star)$. This set satisfies all the constraints of problem $\bs{Q}_1(\bs{x}^{(i)}_\star,\bs{y}^{(i)}_0, \bs{z}^{(i)}_0)$, and therefore is a feasible solution of the optimization problem. As a result, the optimal solution at the $i+1$-th iteration, $t^{(i+1)}_\star$, must satisfy $t^{(i+1)}_\star \leq t^{(i+1)} < t^{(i)}_\star$, which completes the proof of Proposition~\ref{prop 1}.
\balance
\section{Proof of Proposition~\ref{prop 2}}\label{app:2}
Denote $\big( t^{(i)}_\star, \bs{p}^{(i)}_\star, \bs{q}^{(i)}_\star, \bs{x}^{(i)}_\star, y^{(i)}_\star\big)$ as the optimal solution of $\bs{Q}_2(\bs{x}^{(i)}_0,y^{(i)}_0)$ at iteration $i$. We will show that if $x_{\star k}^{(i)} < x^{(i)}_{0k}$ and $y^{(i)}_\star > y^{(i)}_0, \forall k\le K_C$, then by using $x^{(i+1)}_{0k} = x_{\star k}^{(i)}$, $y^{(i+1)}_{0k} = y^{(i)}_{\star k}$ in the $(i+1)$-th iteration, we will have $t^{(i+1)}_\star  < t^{(i)}_\star $. Indeed, by choosing a relatively large initial value $\{x^{(1)}_0\}_{k=1}^{K_C}$ and small value $\{y^{(1)}_{0k}\}_{k=1}^{K_C}$, we always have $x_{\star k}^{(1)} < x^{(1)}_{0k}$ and $y^{(1)}_{\star k} > y^{(1)}_{0k}, \forall k \le K_C$. 

By using $\bs{x}^{(i)}_\star$ at the $(i+1)$-th iteration, we have $x^{(i+1)}_{0k} = x^{(i)}_{\star k}, \forall k$. Therefore, $f(x;x_{\star k}^{(i)})$ is used in the right-hand side of constraint \eqref{eq:Tzf2 c2 app}, where $f(x;a) = e^a(x - a + 1)$ is the first order approximation at $a$ of function $e^x$. Consider a candidate $\bs{x}^{(i+1)} = \{x^{(i+1)}_1, \dots, x^{(i+1)}_K\}$ with $x^{(i+1)}_k \in (\hat{x}_k, x^{(i)}_{\star k})$, where $\hat{x}_k = x_{\star k}^{(i)} - 1 + e^{x^{(i)}_{0k} - x_{\star k}^{(i)}}(x_{\star k}^{(i)} - x^{(i)}_{0k} + 1)$. It is evident that $x^{(i+1)}_k <  x_{\star k}^{(i)}$ and $f(x^{(i+1)}_k;x_{\star k}^{(i)}) > f(x_{\star k}^{(i)}; x^{(i)}_{0k}), \forall k \le K_C$. In addition, consider a candidate $y^{(i+1)} = y^{(i)}_\star + \delta y$, with $\delta y \le \min_{1\le k \le K_C} \{\bar{\mu}_k (x^{(i)}_{\star k} - x^{(i+1)}_k)\}$. Obviously, $f(y^{(i+1)}_k; y^{(i)}_{\star k}) > f(y^{(i)}_{\star k}; y^{(i)}_{0k})$ due to the convexity of $e^y$ function.

Because $f(x^{(i+1)}_k;x_{\star k}^{(i)}) > f(x_{\star k}^{(i)}; x^{(i)}_{0k})$ and $f(y^{(i+1)}_k; y^{(i)}_{\star k}) > f(y^{(i)}_{\star k}; y^{(i)}_{0k}), \forall k \le K_C$, the strict inequality holds in constraints \eqref{eq:Tzf2 c2 app} and \eqref{eq:Tzf2 c3 app}. Thus, there exits $p^{(i+1)}_k > p^{(i)}_{\star k}$ and $t^{(i+1)} < t^{(i)}_\star$ which satisfies constraints \eqref{eq:Tzf1 c1}, \eqref{eq:Tzf2 c2 app} and \eqref{eq:Tzf2 c3 app}. Furthermore, since $\delta y \le \min_{1\le k \le K_C} \{\bar{\mu}_k (x^{(i)}_{\star k} - x^{(i+1)}_k)\}$, constraint \eqref{eq:Tzf2 c1} is also satisfied. Now consider a new candidate set $(t^{(i+1)}, \bs{p}^{(i+1)}, \bs{q}^{(i)}_\star, \bs{x}^{(i+1)}, \bs{y}^{(i+1)})$. This set satisfies all the constraints of problem $\bs{Q}_2(\bs{x}^{(i)}_\star,\bs{y}^{(i)}_\star)$, and therefore is a feasible solution of the optimization problem. As a result, the optimal solution at the $(i+1)$-th iteration, $t^{(i+1)}_\star$, must satisfy $t^{(i+1)}_\star \leq t^{(i+1)} < t^{(i)}_\star$, which completes the proof of Proposition~\ref{prop 2}.

\end{document}